\documentclass[preprint,showpacs,preprintnumbers,amsmath,amssymb,pre,floatfix,]{revtex4}
\usepackage{epsfig} 
\usepackage{graphicx}
\usepackage{dcolumn}
\usepackage{bm}
\usepackage{wasysym/wasysym}

\begin{document}

\title{ Liquid-Liquid Phase Transitions
for Soft-Core\\ Attractive Potentials}

\author{A. Skibinsky$^1$}
\author{S. V. Buldyrev$^1$}
\author{G. Franzese$^{1,2}$}
\author{G. Malescio$^3$}
\author{H. E. Stanley$^1$}

\affiliation{
$^1$Center for Polymer Studies and Department of Physics,
Boston University, Boston, MA 02215, USA\\
$^2$Departament de Fisica Fonamental,
Facultat de Fisica,
Universitat de Barcelona,
Diagonal 647, 08028 Barcelona, Spain\\
$^3$Dipartimento di Fisica, Universit\`a di Messina
and Istituto Nazionale per la Fisica della Materia, I-98166 Messina, Italy
}

\date{EH9063 -- revised: 30 January 2004 -- sbfms.tex}

\begin{abstract}

Using event driven molecular dynamics simulations, we study a three
dimensional one-component system of spherical particles interacting via
a discontinuous potential combining a repulsive square soft core and an
attractive square well.  In the case of a narrow attractive well, it has
been shown that this potential has two metastable gas-liquid critical
points.  Here we systematically investigate how the changes of the
parameters of this potential affect the phase diagram of the system. We
find a broad range of potential parameters for which the system has both
a gas-liquid critical point $C_1$ and a liquid-liquid critical point
$C_2$.  For the liquid-gas critical point we find that the derivatives
of the critical temperature and pressure, with respect to the parameters
of the potential, have the same signs: they are positive for increasing
width of the attractive well and negative for increasing width and
repulsive energy of the soft core. This result resembles the behavior of
the liquid-gas critical point for standard liquids.  In contrast, for
the liquid-liquid critical point the critical pressure decreases as the
critical temperature increases. As a consequence, the liquid-liquid
critical point exists at positive pressures 
only in a finite range of parameters.  We present
a modified van der Waals equation which qualitatively reproduces the
behavior of both critical points within some range of parameters, and
give us insight on the mechanisms  
ruling the dependence of the two critical points on the potential's
parameters. 
The soft core
potential studied here resembles model potentials used for colloids,
proteins, and potentials that have been related to liquid metals,
raising an interesting possibility that a liquid-liquid phase transition
may be present in some systems where it has not yet been observed.

\end{abstract}


\maketitle

\section{Introduction}

The discovery and investigation of liquid-liquid phase transitions in a
one-component system is of current interest, since recent experiments
for phosphorus \cite{KATAYAMA,Monaco} show a first-order phase
transition between two stable liquids in the experimentally accessible
region of the phase diagram.  A liquid-liquid phase transition, ending
in a critical point was initially proposed to explain the anomalous
behavior of network-forming liquids such as H$_2$O
\cite{Debenedetti,HPB,Robinson,VOLGA,pablonewreview,pablogene,%
POOLE1,Angell00,Debene,TANAKA,%
BELLISSENT,ms98,MISHIMA,SOPER,Lee,Guisoni}.  In particular, the density
anomaly, consisting in the expansion under isobaric cooling of these
systems, has been related to the possible existence of a phase
transition between low density liquid (LDL) and high density liquid
(HDL). Simulation results and experimental studies of water predict a
LDL-HDL phase transition in an experimentally inaccessible region of the
phase diagram \cite{POOLE1,TANAKA,MISHIMA,FMS,slt}.  Computer simulations of
realistic models of carbon \cite{Glosli}, phosphorus \cite{MORISHITA},
SiO$_2$ \cite{SAIKA-VOIVOD}, and Si \cite{ANGELL,SHRI_PP} strongly
suggest the existence of first-order LDL-HDL phase transitions in these
substances.  Recently the step changes of the viscosity of liquid metal,
such as Co, have been theoretically interpreted as evidence of
liquid-liquid phase transitions~\cite{Vasin}.
 
The presence of the first-order phase transitions in solids and
solid-solid critical points, determined experimentally~\cite{DZHAVADOV}
and with simulations~\cite{YOUNG,BOLHUIS-SC,STELL2,HEMMER-SC},
have suggested the possibility of the existence of liquid-liquid
critical points and polymorphism in the amorphous 
state~\cite{MITUS,BrazhkinNKPT,Tarjus}. It has been proposed that systems with
solid polymorphism may exhibit several liquid phases with local
structures similar to the local structures of various crystals.
Experimental evidence of sharp structural transitions between liquid
polymorphs of Se, S, Bi, P, I$_2$, Sn, Sb, As$_2$Se$_3$, As$_2$S$_3$,
and Mg$_3$Bi$_2$ are consistent with  
phase diagrams with first-order liquid-liquid
phase transitions~\cite{BrazhkinNKPT,brazh}, 
analogous to the
liquid-liquid phase transition seen in rare earth aluminate 
liquids~\cite{AASLAND,WILDING}.

These results call for a general interpretation of the basic mechanisms
underlying the liquid-liquid phase transition. Here we aim to delineate
the conditions ruling the accessibility of the two liquid phases.
A first step in this direction was taken
in Refs.~\cite{nature,fmsbs}, where we have shown
that a specific isotropic soft-core attractive potential, for a
one-component system, has a phase diagram with LDL-HDL phase transition,
with two fluid-fluid critical points, and with no density anomaly.

Here we extend this analysis by varying the parameters of this potential
(Fig.~\ref{potential2}).  We find that, for a wide range of parameters,
this potential has a phase diagram with a liquid-liquid critical
point, and we show how the phase diagram depends on the parameters.
We develop a modified van der Waals equation (MVDWE) able to describe
the behavior of the two critical points as a function of the potential
parameters, elucidating a mechanism for the liquid-liquid phase
transition and the conditions under which the liquid-liquid critical
point occurs at positive pressure.

In Sec. II we introduce the isotropic soft core potential; in Sec. III
we describe the two different molecular dynamics techniques we use; in Sec. IV
we present our results for different combinations of parameters that
give rise to a liquid-liquid phase transition ending in a
liquid-liquid critical point; in Sec. V we construct a modified van
der Waals equation which can qualitatively reproduce the behavior of
the two critical points; in Sec. VI we discuss the role of potential
parameters in changing the position of the critical points; in
Sec. VII we summarize our results; in Appendix A we present our
simulation results for a simple square well potential.

\section{The isotropic soft-core attractive potential}

For attractive potentials with a sufficiently broad interaction distance,
the phase diagram has a first-order gas-liquid transition ending in a
gas-liquid critical point, and a first-order liquid-solid phase
transition \cite{ELLIOT}.  When the attractive range is small, the
liquid phase and the gas-liquid critical point are metastable with
respect to the solid phase
\cite{LEKKERKERKER,BOLHUIS-SQW,BOLHUIS-SQW2,Frenkel,Stell4}.

For a strictly repulsive soft-core potential, simulations show a phase
diagram with a first-order gas-solid phase transition and a first-order
phase transition between two solids of different densities, but with the
same structural symmetry, ending in a solid-solid critical point
\cite{YOUNG,BOLHUIS-SC,STELL2,HEMMER-SC}.  Recent theoretical
work has suggested that systems with a broad soft-core potential have a
fluid-fluid phase transition and liquid anomalies~\cite{RYZHOV}, or
give rise to stripe phases in two dimensions~\cite{pellicane}.

We have shown in Ref.~\cite{nature} that the combination of a repulsive
soft core with an attractive well is sufficient to give rise to a phase
diagram with two liquid phases. This simple isotropic model potential is
similar to those used in the seminal work of Stell and Hemmer
\cite{STELL}, who studied a soft-core potential in one dimension
(1D). Similar potentials were studied in 2D and 3D showing phase
diagrams with a possible liquid-liquid critical point
\cite{jagla,REZA}.

The 3D isotropic potential we consider (Fig.~\ref{potential2}) has a
hard core (infinite repulsion) at distance $a$, a repulsive soft core of
width $w_R$ and energy $U_R>0$, and an attractive square well of width
$w_A$ and energy $-U_A<0$~\cite{nature,fmsbs}.  The potential has three
parameters: $w_R/a$, $w_A/a$, and $U_R/U_A$, where $a$ and $U_A$ have
been chosen as units of length and energy, respectively.  Though this
potential is discontinuous, it is similar to model potentials for
complex fluids, such as colloids, protein solutions, star polymers
\cite{Frenkel,AUER,STILLINGER, STILLINGER2,colloids1,colloids,LoVerso},
and resembles pair potentials proposed for water~\cite{STILLINGER}, or
that have been related to liquid metals under specific conditions
~\cite{debene2,WAX,Li}. 

This potential with parameters $w_R/a=1.0$, $w_A/a=0.2$, and
$U_R/U_A=0.5$ has a phase diagram with gas-LDL and gas-HDL first-order
phase transitions, each ending in a critical point in the supercooled
fluid region~\cite{nature}.  Both liquid phases are metastable with
respect to a single crystal phase and no density anomaly is
observed~\cite{fmsbs}.

In this paper we present systematical MD studies of the phase diagrams
for this potential (Fig.~\ref{potential2}). 
By varying the parameters of the potential, $w_{A}/a$, $w_{R}/a$, and
$U_{R}/U_A$, we relate the attractive and repulsive
components of the potential to the appearance and stability of the
liquid-liquid phase transition and critical points. 

\section{Molecular dynamics simulations}

We perform MD simulations of $N=850$ particles of unit mass $m$ at
constant volume $V$, and constant temperature $T$,
interacting via the potential described above (Fig.~1).  The details of
the event-driven MD we use are presented in Refs.~\cite{nature,fmsbs}. 
We measure time in units of $aU_A^{-1/2}m^{1/2}$ and record potential
energy and pressure every $\Delta t=100$ time units.  To understand the
effect of each parameter on the phase diagram of our system, we simulate
sixteen sets of potential parameters (Table~\ref{sets}).  After a
preliminary screening, we choose to study the region of parameter space
where the low-density, gas-liquid critical point $C_1$ always has a
critical temperature above that of the high-density critical point
$C_2$. Therefore, while in Refs.~\cite{nature,fmsbs} $C_2$ is a gas-HDL
critical point, here $C_2$ is a LDL-HDL critical point. 
As shown in in Refs.~\cite{nature,fmsbs}, $C_2$ can lie in the 
supercooled metastable phase, close to the line of
homogeneous nucleation, as in 
water or silica~\cite{ms98,FMS,SHRI_PP}. We make
certain that all our calculations are
performed before the onset of crystallization, as discussed in
Ref.~\cite{fmsbs}. The description of the crystal phases goes beyond
the goal of this work. To optimize our
analysis we use two different MD methods.

\subsection{Isothermic method}

The first method is a straightforward calculation of the phase diagram's
state points. For each state point with given $\rho = N/V$ and $T$, we
perform typically ten independent simulations of $t\approx 2\times 10^3$
time units.  We estimate the error in pressure measurements from the
standard deviation of the ten averaged values computed for each
independent simulation. The state points along the isotherms are
approximated by a two-variable polynomial
$P(\rho,T)=\sum_{i\kappa}a_{i\kappa}\rho^iT^\kappa$ obtained by the
least squared fit of all the state points in the vicinity of the
critical point. This fitting implies mean field critical
exponents~\cite{mf} and may produce incorrect results in the close
vicinity of the critical point. However, this method helps us fit the
state points, known with statistical errors, by approximate polynomial
isotherms and thus obtain the approximate position of the critical point.

The coexistence curves are calculated using Maxwell's equal area
construction and spinodal line is estimated by locating the maxima and
minima of the isotherms. After calculating the state points, isotherms,
coexistence curves, and spinodal lines, we estimate the critical pressure,
temperature, and density for $C_1$ and $C_2$ ($P_{C_1}$, $T_{C_1}$,
$\rho_{C_1}$, $P_{C_2}$, $T_{C_2}$, and $\rho_{C_2}$, respectively) as
the point where coexistence and spinodal curves meet, coinciding at
their maxima.  We apply this method to six sets of potential parameters
(ii, iii, iv, xi, xii and xiv in Table~\ref{sets}). The results are
presented in Figs.~2--7 in the pressure-density ($P$-$\rho$) phase
diagrams. The estimates of the critical points are presented in
Table~\ref{new}.

\subsection{Isochoric method}

The isothermic method gives us fairly complete information about the details of
the phase diagrams, but requires much computation to calculate enough
state points for accurate isotherms. Thus, in order to find the
positions of critical points for a wide range of potential parameters,
we adopt a faster but less accurate MD method.
For sets of parameters close to the sets of parameters studied with the
isothermic method, we estimate the location of the spinodal line by
evaluating the intersections of isochores in the $P-T$ plane. We first
equilibrate several configurations at a high initial temperature
$k_BT_I/U_A=2.0$ for several values of density above and below the
densities where we expect to find $\rho_{C_1}$ and $\rho_{C_2}$. At
constant density, the system is slowly cooled down from $T_I$ to a final
temperature $k_BT_F/U_A=0.1$ during a simulation time of $10^4$ time
units~\cite{ENERGY}.

The average values of $T$ and $P$ are recorded each 100 time units,
which is comparable to the equilibration time of the system for $k_B
T/U_A>0.5$.  As the temperature decreases, the equilibration time
increases and the method becomes less reliable. Thus, we use this method
to estimate pressure and potential energy for $k_B T/U_A>0.5$.

The error bars of each measurement are of the order of the non-monotonic
jumps of the isochores (see Fig.~\ref{isochores} inset). The
intersection is determined by fitting isochores with smooth polynomial
fits. The best results can be achieved by quadratic fits in the
temperature range including the region of possible isochore crossing
extending from $0.9T_{C}$ to $1.5T_{C}$, so that the tentative critical
temperature $T_C$ is inside this interval.

Since at the spinodal line $(\partial P/\partial \rho)_T=0$, two
isochores with two close values of density must intersect in the
vicinity of the spinodal line.  By definition, the critical point
corresponds to the maximum temperature on the spinodal. Therefore, the
critical pressure and temperature can be evaluated by estimating the
pressure corresponding to the maximum temperature at which isochores
intersect. The critical density can be estimated as ($\rho_1+
\rho_2)/2$, where $\rho_1$ and $\rho_2$ are the densities of the
two isochores intersecting at the highest temperature
(Fig.~\ref{isochores}). The critical point values estimated with this
method are presented in Table~\ref{CPs}.

This approximate method is allowed as long as we use it to estimate
the critical points of potentials with sets of parameters close to
those for which we have done a detailed study using the isothermic
method.  We apply the isochoric method to 16 sets of the potential
parameters.  The comparison of the two methods
(Tables~\ref{new}-\ref{CPs}) shows that the resulting estimates of
critical $P$, $T$, and $\rho$ of $C_1$ and $C_2$ are consistent.

\section{Phase Diagram Results}

Our results in Figs.~\ref{md1}--\ref{md6} clearly show that the phase
diagram strongly depends on the potential parameters. For example, phase
diagrams in Figs.~\ref{md1}--\ref{md3} have fluid phases (gas, LDL,
and HDL) at positive pressures, while for the phase diagrams in
Figs.~\ref{md4}--\ref{md6}, the high-density critical point appears at
negative pressures, i.e. in the region of stretched fluid.

To investigate how the position of critical points depends on the
potential parameters, we vary one of the three parameters $w_A/a$,
$w_R/a$, $U_R/a$ at a time, keeping the other two constant. The behavior
of $T$, $P$, and $\rho$ for $C_1$ and $C_2$ (Fig.~\ref{all} and
Table~\ref{summ-table}) are presented in the following.

\subsection{Effect of the square-well width $w_A$}

By keeping $w_R/a=0.5$ and $U_R/U_A=2.0$ constant, we find
(Fig.~\ref{all}a--c, Fig.~\ref{wl_w12.gc}) that by increasing well
width $w_{A}$, $\rho_{C_1}$ is almost unaffected, while $\rho_{C_2}$
decreases, $T_{C_1}$ and $T_{C_2}$ increase, $P_{C_1}$ increases,
while $P_{C_2}$ decreases. For $w_A/a>0.7$ the LDL-HDL critical point
$C_2$ occurs at negative pressures, as in Fig.~\ref{md4}.  Hence,
$C_2$ lies in the stretched fluid region and, therefore, it is
metastable. In order to have a stable LDL-HDL critical point, the
attractive distance $w_{A}/a$ must be sufficiently narrow, so that
$C_2$ occurs at positive pressures. A too narrow well, however,
enhances crystallization
\cite{fmsbs,LEKKERKERKER,BOLHUIS-SQW,BOLHUIS-SQW2,Frenkel,Stell4}
so that the high-density critical point
shifts below the line of spontaneous crystallization, becoming
difficult to observe. Thus the
liquid-liquid critical point is observable in our MD simulations 
only for intermediate values of
$w_A/a$.

\subsection{Effect of the shoulder width $w_R$}

Increasing the width of the repulsive interaction $w_{R}$, while keeping
$w_A/a=0.7$ and $U_R/U_A=2.0$ constant, we find (Fig.~\ref{all}d--f,
Fig.~\ref{sh_w12.gc}) that both $\rho_{C_1}$ and $\rho_{C_2}$ decrease,
$T_{C_1}$ decreases, while $T_{C_2}$ increases, and both $P_{C_1}$ and
$P_{C_2}$ decrease. For $w_{R}/a < 0.4$ the dynamics of the system in
the vicinity of the expected high density critical temperature become
too slow and the equilibration time becomes too long, with respect to
our simulation time, to measure the equilibrium state points with
sufficient accuracy. Furthermore, as expected for decreasing $w_{R}$,
$T_{C_2}$ approaches $T=0$ (Fig.~\ref{all}e), suggesting that $C_2$
disappears for $w_R/a=0$. At $w_{R}/a > 1.0$ the system spontaneously
crystallizes at high density without showing a second critical point
$C_2$. Hence, the width of the shoulder $w_{R}/a$ must be of an
intermediate value for $C_2$ to be observed above the lines of spontaneous
crystallization and outside the region of very slow dynamics, at least for
our choice of $w_A$ and $U_R$.

\subsection{Effect of the shoulder height $U_R$}

For $w_{R}/a=0.5$ and $w_{A}/a=0.9$, we increase the repulsive energy
$U_R$ and find (Fig.~\ref{all}g-i, Fig.~\ref{sh_h12.gc}) that for
increasing $U_R$, $\rho_{C_1}$ decreases, while $\rho_{C_2}$ is almost
unaffected, both $T_{C_1}$ and $T_{C_2}$ decrease, $P_{C_1}$ decreases,
while $P_{C_2}$ rapidly increases.  For $U_{R}/U_A< 2.0$ the
high-density phase transition occurs at very low negative pressures and
the fluid phases are highly metastable.  For $U_{R}/U_A > 4.0$ the
diffusion in the system in the vicinity of the high density critical
point becomes markedly slow, due to the soft core becoming less
penetrable and assuming the role of an effective hard core.  Therefore,
an intermediate repulsive energy is needed to observe $C_2$ in our MD
simulations.

\section{Modified van der Waals equation}

To rationalize the dependence of the temperature, pressure and density
of the two critical points on the potential's parameters, we 
develope a simple mean field theory that gives rise
to a modified van der Waals equation (MVDWE), 
\begin{equation}
P=\frac{\rho T}{1-\rho B(\rho,T)}-A\rho^2 ,
\label{MVDWE}
\end{equation}
that has the same form of the standard van der Waals equation (see
Appendix A), but with an excluded volume $B(\rho,T)$ depending on the
density and temperature of the state point and increasing with $w_R/a$,
and with a strength of attraction $A$ that increases with $w_A/a$ 
and decreases with $U_R/U_A$. It should be pointed out that a 
different modification of the van der
Waals equation \cite{Tejero} also gives rise to the
high density critical point. In contrast with our work, 
Ref. \cite{Tejero} is particularly suitable
for density dependent potentials since it assumes a constant
excluded volume $B$ and a density dependent attractive term $A(\rho)$. 

For a system with a hard core and a soft core, one can assume that the
effective excluded volume $B(\rho,T)$ changes with temperature and
density~\cite{Stishov}. Indeed, at low densities and low temperatures, particle
cannot penetrate into the soft core so $B(\rho,T)\approx B_2$ where
$B_2=2{\pi}(a+w_R)^3/3$ is the excluded volume associated with the soft core.
In contrast, for high densities and high temperatures, particles easily
penetrate into the soft core and $B(\rho,T)\approx B_1$, where
$B_1=2{\pi}a^3/3$ is the excluded volume associated with the
hard core. More specifically, $B(\rho,T)$ must be an analytical function of
its parameters such that $\partial{B(\rho,T)}/\partial{T}<0,$
$\partial{B(\rho,T)}/\partial{\rho}<0$,
\begin{equation}
\lim_{T\to\infty}B(\rho,T)=B_1,
\end{equation}
and
\begin{equation}
\lim_{T\to 0}B(\rho,T)=\left\{
\begin{array}{ll}
B_2,    & \rho \leq 1/B_2 \\
1/\rho, & 1/B_1 >\rho> 1/B_2,
\end{array}\right.
\end{equation}
from which it follows that $B(\rho,T)<1/\rho$ for any $\rho$ and $T>0$. 

Since in any case van der Waals equation can
give us only qualitative agreement with reality, we can select any model
function $B(\rho,T)$ which satisfies the above conditions. Never the
less, it is desirable to select $B(\rho,T)$ in such a way that it will
describe the behavior of some physical system for which the analytical
solution can be found. One dimensional system of particles with a pair
potential
\begin{equation}
U(r)=\left\{
\begin{array}{ll}
\infty, & r < B_1 \\
U_R,    & B_1\leq r < B_2 \\
0,      & r\geq B_2 ~~~ ,
\end{array}\right.
\end{equation}
provides such a solution. Applying the Takahashi method~\cite{REZA,Lieb},
we obtain the Gibbs potential
\begin{equation}
G=-k_BTN_1 \ln(\Psi k_BT/B_1P_1)
\end{equation}
where $N_1$ is the number of particles, $T$ is temperature, $P_1$ is
pressure of the one-dimensional system and 
\begin{equation}
\Psi(T,P_1)=(e^{-P_1B_1/k_BT} -e^{-P_1B_2/k_BT})e^{-U_R/k_BT} +
 e^{-P_1B_2/k_BT}. 
\end{equation}
Accordingly $V_1=\partial{G}/\partial{P_1}$ and
$S_1=-\partial{G}/\partial{T}$ are the volume and entropy of the
one-dimensional system, and $U_1=G-P_1V_1+TS_1$ is the potential energy
for the one dimensional system. The fraction of the soft
cores $f(\rho,T)$ penetrated by the particles is
$f(\rho,T)=U_1(P_1,T)/(N_1U_R)$ where $P_1$ must be determined as a
function of $\rho$ from the equation
$\partial{G}/\partial{P_1}(P_1,T)=V_1\equiv N_1/\rho$. The value
$f_{\infty}\equiv f(\rho,\infty)$ is the fraction of the soft cores
penetrated by the particles in the high-temperature limit in which soft
cores play no role. It can be computed assuming a Poisson distribution
of interparticle distances:
$f_{\infty}=1-e^{(B_1-B_2)/(1/\rho-B_2)}$. The probability that the soft
core does not reflect the neighboring particle is equal to the fraction
of these two quantities $f/f_{\infty}<1$. In this case, the excluded
volume is equal to $B_1$. In the opposite case with probability
$1-f/f_\infty$, the excluded volume is equal to $B_2$. Hence, the
effective excluded volume
\begin{equation}
B(\rho,T)=f/f_{\infty}B_1+(1-f/f_{\infty})B_2,
\end{equation}
where, 
\begin{equation}
\frac{f}{f_{\infty}}=\frac
{(e^{-P_1B_1/k_BT}-e^{-P_1B_2/k_BT}) e^{-U_R/k_BT}}
{\Psi(P_1,T)
(1-e^{(B_1-B_2)/(1/\rho-B_1)})}~~~,
\end{equation}
and $P_1$ must be found from the equation 
\begin{equation}
\frac{1}{\rho}=\frac{k_BT}{P_1}+
\frac{(B_1e^{-P_1B_1/k_BT}-B_2e^{-P_1B_2/k_BT})e^{-U_R/k_BT}+B_2e^{-P_1B_2/k_BT}}
{\Psi(P_1,T)}.
\end{equation}

Figure~\ref{vander}a illustrates the behavior of $B(\rho,T)$ for a
particular set of parameters. It is clear that $B(\rho,T)$ satisfies
all the physical conditions we impose on the effective excluded volume. The
modified van der Waals equation~({\ref{MVDWE})
has two critical points: one for low density $\rho <<1/B_2$ and
another for high density $\rho \approx 1/B_2$, whose positions on the
phase diagram of the dimensionless variables $\tilde T=k_BT/U_R$,
$\tilde P=B_1P/U_R$, and $\tilde \rho=B_1\rho$ depend on the
dimensionless parameters of the MDVWE: $B_2/B_1$ and $A/(U_R B_1)$. 
Figure~\ref{vander}b shows a $P-T$ diagram with two critical
points $C_1$, $C_2$, for a particular set of parameters, for which the
position of the critical points are similar to the positions found in
our simulations, i.e. $T_{C_2}<T_{C_1}$.
 
Now we can relate the parameters of the Eq.~\ref{MVDWE}
to the potential parameters used in our 
simulations.  The parameters 
$B_1$ and $B_2$ are increasing functions of the
hard core diameter $a$ and the shoulder width $w_R$, respectively. The
parameter $U_R$ has an identical meaning in MVDWE and in simulations. 
The strength of attraction $A$ is an increasing function of
$w_A$ and a decreasing function of $U_R$. 
Indeed, according to the formula of the second virial
coefficient $v_2$ for our potential, we have~\cite{Huang63}
\begin{equation}
v_2 = B_1 + (1-e^{-U_R/k_BT})(B_2-B_1) +
(1-e^{U_A/k_BT})\left[\frac{2\pi}{3}(a+w_R+w_A)^3-B_2\right] ~~~~.
\end{equation}
For large $T$, it has the form $v_2=B-A/k_BT +O(T^{-2})$
with $A=U_Av_A -U_Rv_R$ where $v_A$ and $v_R$ are positive quantities
with the dimension of a volume depending on $a$, $w_R$, and $w_A$, 
$B=\lim_{T \to \infty} v_2$, and  $A=\lim_{T  \to \infty}T(B-v_2)$.  
Hence, in this limit, the  
virial expansion $P=k_BT\rho + k_BTv_2\rho^2 +O(\rho^3)=
k_BT\rho(1+B\rho)-A\rho^2 + O(\rho^3)$ 
can be rewritten in the form
of the van der Waals equation $P=k_BT\rho/(1-B\rho)-A\rho^2 +O(\rho^3)$.

From the equations above we can derive the functional relation for
$v_A$ and $v_R$ in the limit $T\to \infty$, that are
$v_A=(2\pi/3)(a+w_R+w_A)^3-B_2$ and $v_R=B_2-B_1$. By using these
relations it is possible to see that in general $A$ is an
increasing function of $w_A$ and a decreasing function of $U_R$. 
The derivative $\partial A/\partial w_R$ may have different sign depending
on other parameters. Although at finite $T$ these relations could
be valid only to the leading order, it is reasonable to assume that
$A$ increases with $w_A$ and decreases with $U_R$ at any $T$.

However, to simplify our qualitative study of the MVDWE, we assume the
parameters $A$, $B_2$ and $U_R$ are independent.  By varying these
parameters one at a time and by relating $B_2$ to $w_R$, and $A$ only
to $w_A$, we found that the MVDWE predicts that the derivatives of the
low-density critical point values, $T_{C_1}$, $P_{C_1}$, $\rho_{C_1}$,
with respect to each of the parameters of MVDWE, have the same sign
and this sign is negative for $B_2(w_R)$ and $U_R$, which increase
repulsion, and this sign is positive for $A(w_A)$, which increases
attraction. These results are consistent with the MD results for the
low-density critical point.  For the high-density critical point
values the MVDWE predicts for some parameters non-monotonic behaviors
and we find that there are regions of parameters where the critical
values as a function of $A$, $B$ and $U_R$ have the same qualitative
behaviors as those found in the simulations as a function of $w_A$,
$w_R$ and $U_R$, respectively (Fig.~\ref{new3x3}). Specifically, we
observe monotonic behaviors of $\rho_{C_2}(B_2)$, $T_{C_2}(A)$,
$T_{C_2}(B_2)$, and $P_{C_2}(U_R)$ which qualitatively coincide with the
corresponding behaviors in simulations (Figs \ref{all} and
\ref{new3x3}~d,b,e, and i).  We find non-monotonic behaviors of
$\rho_{C_2}(A)$, $\rho_{C_2}(U_R)$, $T_{C_2}(U_R)$, $P_{C_2}(A)$, and
$P_{C_2}(B_2)$, which qualitatively coincide with the corresponding
behaviors in simulations in certain range of parameters (Figs
\ref{all} and \ref{new3x3}~a,g,h,c, and f).

These
observations indicate that the behavior of the critical points in
simulation may also become non-monotonic in the range of parameters
that we do not explore. For example, $T_{C_2}(U_R)$ may start to
increase for large $U_R/U_A>4$ and small $w_R/a<0.5$. Another
interesting prediction of the MVDWE is that for large
$B_2/B_1>B_T(AB_1/U_R)$, where $B_T(x)$ increases from $B_T(0.7)=1$ to
$B_T(3.2)=1.7$ the high-density critical temperature becomes larger
than the low-density critical temperature as in simulations of Refs.
\cite{nature,fmsbs}, for which the repulsive shoulder $w_R/a=1$  was
much wider than the attractive well $w_A/a=0.2$. Also, MVDWE predicts
the existence of the third, very high density critical point for large
$B_2/B_1$ and large $AB_1/U_R$, which was recently observed 
in simulations with a wide soft core \cite{bs}.

\section{Role of Potential Parameters}

In the following we will present
the comparison between the MD results and MVDWE predictions.

\subsection{The low density critical point}

First we note that at low densities,
corresponding to the critical point $C_1$, and at sufficiently low
temperatures, particles do not penetrate into the repulsive region,
$r<a+w_R$. Therefore, we can assume that, at low enough temperatures and
densities, the system is interacting via an effective potential given
by a simple square-well with hard core $a+w_R$,
an attractive well of relative width $w_A/(a+w_R)$ and attractive
energy $U_A$.  

Indeed, for increasing width of the attractive well $w_A$, $\rho_{C_1}$ is
roughly constant (Fig.~\ref{all}a) and $T_{C_1}$ and
$P_{C_1}$ increase (Fig.~\ref{all}b and c). 
This behavior is consistent with the predictions of the standard van
der Waals theory for the gas-liquid critical point for a square-well
potential (see Appendix A), that 
yields Eqs.~(\ref{equ12}--\ref{equ14}).
This result supports the idea that the effect of the
soft core is negligible at low densities. 
The MVDWE
also predicts strong increase of
$P_{C_1}$ and $T_{C_1}$ with the strength of attraction $A$,
which increases with $w_A$. For $\rho_{C_1}$, the MVDWE predicts a weak
increase, which can be observed in Fig.~\ref{all}a for
larger $w_A$.

For increasing width of the repulsive shoulder $w_R$, $\rho_{C_1}$
decreases and saturates (Fig.~\ref{all}d) and $T_{C_1}$ and
$P_{C_1}$ decrease (Fig.~\ref{all}e and f). 
As a consequence of the above considerations, the
increase of $w_R$ corresponds to a decrease of this effective attractive
parameter.  Accordingly, $T_{C_1}$ and $P_{C_1}$ display
the behavior predicted for the decrease of the attractive width in van
der Waals theory for the square-well potential
(Eqs.~\ref{equ13}-\ref{equ14}).  Moreover, the behavior of $\rho_{C_1}$
is consistent with Eq.~(\ref{equ12}), which predicts $\rho_{C_1}\sim
1/a^3$. Indeed, in our case, the hard core is replaced by the effective
hard core, hence $a^3\rho_{C_1}\sim a^3/(a+w_R)^3$, which is a
decreasing function of $w_R$. The calculations using MVDWE completely
confirms these predictions by showing 
that $T_{C_1}, P_{C_1}$, and
$\rho_{C_1}$ all decrease with increasing $w_R$ (or 
$B_2=2\pi(a+w_R)^3/3$ as in Sec.V).

For increasing repulsive energy $U_{R}$, the behaviors of $\rho_{C_1}$,
$T_{C_1}$ and $P_{C_1}$ are the same as those observed for increasing
$w_R$ (Fig.~\ref{all}g--i).  This can be understood by considering that
the increase of $U_R$ effectively decreases the penetrability of
interparticle distances $r<a+w_{R}$. The soft core becomes an effective
hard core as described above. In particular, the saturation of
$\rho_{C_1}$, already observed in Fig.~\ref{all}d, is now more evident
and an analogous behavior is now also seen for $T_{C_1}$ and
$P_{C_1}$. This result shows that for a high-enough repulsive energy
$U_R$ and low-enough $T$, the soft-core potential is equivalent to a
hard-core potential, for which there is no dependence of the critical point
on $U_R$. Again, the MVDWE agrees with these predictions by showing that
$T_{C_1}, P_{C_1}$, and $\rho_{C_1}$ decrease with increasing $U_R$. 

\subsection{The high density critical point}

For increasing $w_A$, the critical point density $\rho_{C_2}$
decreases, the temperature $T_{C_2}$ increases, and the pressure
$P_{C_2}$ decreases (Fig.~\ref{all}a--c). This finding is in 
agreement with MVDWE predictions for a wide range of parameters.
(Fig.~\ref{new3x3}a--c). 
The behavior of
$T_{C_2}$ is consistent with the idea that the increase of the
attractive distance increases the overall attractive strength of the
potential, allowing more particles to fit within the attractive
interaction range. As a consequence, the system enters the low-energy
and high-density fluid phase at a higher temperature, i.e. $T_{C_2}$
increases.  Hence, the increase of $w_A$ increases the average kinetic
energy of particles at $C_2$, favoring the overcoming of the soft-core
shoulder at low pressures, and we can expect that $P_{C_2}$ decreases.
Moreover, the increase of $w_A$ decreases
the number of elastic interparticle collisions at the
soft-core distance, hence decreases their contribution to the virial
expression for the pressure \cite{books} (see Eq.~(11) in
Ref.\cite{fmsbs}), decreasing the critical pressure $P_{C_2}$.  Note
that the behavior of $P_{C_2}$ in this case is the opposite of the
behavior for $P_{C_1}$ (Fig.~\ref{all}c). With the
decrease of pressure, the density must also decrease, so we can
conclude that $\rho_{C_2}$ must decrease with increasing $w_A$, in
agreement with our simulation results.

For increasing $w_R$, $T_{C_2}$ increases and saturates, and both
$P_{C_2}$ and $\rho_{C_2}$ decrease with a tendency toward saturation
(Fig.~\ref{all}d--f), in agreement with predictions of
MVDWE for a wide range of parameters (Fig.~\ref{new3x3}d--f). 
This happens because the
transition from the LDL to the HDL is characterized by the penetration
of particles into the repulsive soft cores of their
neighbors. Therefore the repulsive soft-core distance $a+w_R$
characterizes the typical distance between the particles at
$C_2$. Hence the increase of $w_R$ reduces the critical density
$\rho_{C_2}$. The behavior of pressure follows the behavior of
density, as in the case of $w_A$, while the derivatives of pressure and
temperature must have the opposite sign due to the same arguments as
above.

For increasing $U_R$, $P_{C_2}$ increases, $\rho_{C_2}$ slowly
increases, and $T_{C_2}$ decreases (Fig.~\ref{all}g--i).  The
predictions of MVDWE coincide with the behavior of $P_{C_2}$ and
$\rho_{C_2}$ in a wide range of parameters (Fig.~\ref{new3x3}g,i)  . 
However, the theory apparently predicts an increasing $T_{C_2}$ 
with $U_R$, except  for very small $U_RB_1/A<0.4$ and large $B_2/B_1>1.5$
(Fig.~\ref{new3x3}h). This discrepancy arises from the fact
that, although it is physically clear that the attractive strength $A$
is a decreasing function of $U_R$, we find the explicit
dependence of $A$ on $U_R$ only in the limiting case $T\rightarrow
\infty$, while for finite $T$ we assume them to be independent.
Hence, we ignore that an increase of $U_R$ decreases $A$ which induces,
as shown in Sec.V, a decrease in $T_{C_2}$.

The behavior of $P_{C_2}$ is easier to understand. Indeed, the
pressure at which the repulsive shoulder can be overcome increases
with $U_R$, which is consistent with the increase of $P_{C_2}$ with
$U_R$. This effect is expected to be more evident at high $U_R$ and to
saturate for decreasing $U_R$, which is consistent with our
results. The critical density $\rho_{C_2}$ increases with $U_R$, for
small values of $U_R/U_A$, as a consequence of the increase of
$P_{C_2}$, and is practically independent of $U_{R}$ when the soft
core plays the role of an effective hard core, i. e. for large enough
$U_R/U_A$. The decrease of $T_{C_2}$ with the increase of $U_R$ is
more difficult to explain. 
Nevertheless, the same argument as in the case
of $w_A$ and $w_R$ which predicts that the derivatives $T_{C_2}$ and
$P_{C_2}$ must have the opposite signs may apply in this case as well.
Finally, we note that increasing with $U_R$ and decreasing with $w_R$, the
behavior of $P_{C_2}$ in three dimensions is consistent with its behavior in 
the one dimensional case, for which
$P_{C_2}/U_A = (U_{R}/U_A + 1) / w_{R}$ \cite{PhysABig}. 

\section{Discussion and conclusions}

We have studied an isotropic attractive 
soft-core square potential in three dimensions
that has a phase diagram with a gas-liquid critical point
$C_1$ and a liquid-liquid critical point $C_2$, separating 
a HDL and a LDL phases.
We have investigated,
with molecular dynamics simulations, how the critical
density, temperature and pressure of the two critical points vary as a
function of the three parameters of the potential, that are the repulsive
energy $U_R/U_A$ in units of the attractive energy $U_A$, the repulsive width
$w_R/a$ in units of the hard core $a$, and the attractive width
$w_A/a$ in units of $a$.

Table~\ref{summ-table} and Fig.~\ref{all} 
show our results for the $\rho$, $T$, and
$P$ of $C_1$ and $C_2$ for varying parameters of the potential.  
To summarize, the behavior of $C_1$ is consistent with that of a
system interacting via an effective square-well potential, with
a hard core $a+w_R$, a relative attractive well
$w_A/(a+w_R)$ and attractive energy $U_A$. The increase of
$U_R/U_A$ or of $w_R/a$ decreases the effective attractive strength
and this effect saturates for large values of $U_R/U_A$. 

This behavior is perfectly predicted by the simple mean field MVDWE. 
In MVDWE, as in the standard
van der Waals equation, the relevant physical parameters are 
the excluded volume $B$ and the stength of attraction $A$, but now
we assume that $B=B(\rho,T)$ increases with $w_R/a$ and 
depends on the state point, while $A$ increases with $w_A/a$ and
decreases with $U_R/U_A$. 

The MVDWE predictions are consistent also with our results on $C_2$.
In general, 
this approach rationalizes why 
the increase of $U_R/U_A$ and the decrease of
$w_A/a$ have the same qualitative effect on the critical points, since 
they have the same effect of the attractive strength.
Moreover, it rationalizes the effect of increasing $w_R$ via the
increase of the excluded volume, hence the decrease of the critical
densities. This decrease induces a decrease of the critical pressures 
$P_{C_1}$ and $P_{C_2}$, as a consequence of the mechanical stability
of the fluid phases.

While for the
low-density critical point $C_1$, the decrease of $P_{C_1}$
occurs with the decrease of $T_{C_1}$, 
i.e. their derivative 
with respect to the potential parameters always have the
same sign, for the high-density critical point $C_2$ the critical
pressure and temperatures always have derivatives with
the opposite sign. 
This behavior can be understood in terms of the
number of elastic collisions with the soft core, which decreases as
$T_{C_2}$ increases, reducing the virial contribution to the pressure.
At the same time, the increase of $T_{C_2}$ reduces the pressure
$P_{C_2}$ necessary to overcome the repulsive soft core and enter into
the HDL phase.

As a consequence, the high density critical point $C_2$ exists at positive
pressure only in a finite region in the parameter space. Indeed, when
the attraction is too strong, i.e. $w_A/a$ is too large or
$U_R/U_A$ is too small, the pressure $P_{C_2}$ becomes negative.
On the other hand, when the strength of attraction is too week, $C_2$
occurs in the deeply supercooled liquid phase, becoming difficult to
observe as in the experimental situation of
water or silica~\cite{ms98,FMS,SHRI_PP}. 

In conclusion, the behavior of both low-density and high-density
critical points qualitatively obeys the mean field predictions of the
modified van der Waals theory based on effective excluded volume
which varies between the hard-core value for high temperature and
low density and the soft-core value for low temperature and low density.  The
quantitative theory based on the thermodynamic perturbation
approximations or various integral equation closures~\cite{lowen} is
yet to be developed.

Confirming the results presented in Refs.~\cite{nature,fmsbs}
we do not find density anomaly.
Our simulations show that density anomaly is
unlikely to exist for the discontinuous double-step potentials shown
in Fig.~\ref{potential2}, in contrast to ramp potentials~\cite{jagla}
and a Gaussian soft core potential~\cite{STILLINGER2}.

Our results may be relevant for experiments on systems that can be
described by an isotropic soft-core attractive potential and have no
density anomaly, such as colloids, protein solutions or liquid
metals. Indeed, our results show that in these systems the possibility
of the existence a liquid-liquid phase transition will depend on the
relative ratio between the attractive and the repulsive part.

\subsubsection*{Acknowledgments}

We thank C. A. Angell, V. V. Brazhkin, A. Geiger, I. Kohl, T. Keyes,
T. Loerting, V. Ryzhov, R. Pastor, and F. W. Starr for helpful
discussions and the NSF Chemistry Division (CHE-0096892) for support.

\appendix
\section{Square-well Fluid System}
\label{sqwell}

Here we recall the case of a square well attractive potential, where the
only parameter is the width of the attractive well $w_A/a$, since the
hard core distance $a$ and the attractive energy $U_A$ can be taken as
units of distance and energy, respectively.  In particular, we show how
the gas-liquid critical point density $\rho_c$, temperature $T_c$, and
pressure $P_c$ depend on $w_A/a$.

Even in this simple case, the phase diagram has no exact analytical
solution and one must rely on various approximations and numerical
simulations~\cite{ELLIOT,lowen,YOUNG-SQW,vega,acedo}. Using MD
simulations of $N = 850$ particles, we verify that the behavior of
$\rho_c$, $T_c$, and $P_c$, are approximately linear for a wide range
of $w_A/a$ (Fig.~\ref{fit-single-well},
Table~\ref{fit-parameters-single-well}) \cite{ELLIOT}. The values of
$a^3\rho_c$ decreases for increasing $w_A/a$ and $a^3P_c/U_A$ and
$k_BT_c/U_A$ increase with $w_A/a$.

Except for density, these results are in agreement with the van der
Waals theory (see, e.g., Ref.~\cite{Huang63}).  The equation of state in
the van der Waals theory is given by
\begin{equation}
\label{vanderW}
 P=\frac{k_B T \rho}{1-\rho B} - A\rho^2,
\end{equation}
where $B=\frac{2}{3}a^3\pi$ has the meaning of excluded volume per
particle and $A$ is a quantity, with the dimension of the product of
energy and volume, characterizing the strength of attraction between
particles. Therefore, $A$ can be related to the product of $U_A$ and the
volume of the attractive well, which is proportional to $w_Aa^2$ for
small $w_A/a$.

The position of a critical point must satisfy the equations
\begin{equation}
\label{equ1}
\left. \frac{\partial P}{\partial \rho} \right|_{T_c} = 0 ~,
\end{equation}
\begin{equation}
\label{equ2}
\left. \frac{\partial^{2} P}{\partial \rho^{2}}\right|_{T_c} = 0 ~.
\end{equation}

For the van der Waals equation of state Eq.~(\ref{vanderW}),
Eqs.~(\ref{equ1}) and~(\ref{equ2}) yield the coordinates of the critical
point:
\begin{equation}
\label{equ9}
\rho_c=\frac{1}{3B} ~,
\end{equation}
\begin{equation}
\label{equ10}
k_BT_c=\frac{8}{27}\frac{A}{B} ~,
\end{equation}
\begin{equation}
\label{equ11}
P_c=\frac{1}{27}\frac{A}{B^2} ~.
\end{equation}
Hence:
\begin{equation}
\label{equ12}
\rho_ca^3 \sim const ~,
\end{equation}
\begin{equation}
\label{equ13}
\frac{k_BT_c}{U_A} \sim \frac{w_a}{a} ~,
\end{equation}
\begin{equation}
\label{equ14}
\frac{a^3P_c}{U_A} \sim \frac{w_a}{a} ~,
\end{equation}
which predict that $k_BT_c/U_A$ and $a^3P_c/U_A$ increase with
$w_A/a$, while $\rho_ca^3$ does not depend on it.

\break\newpage

\begin{table}[hp]
\caption{\label{sets} Sets of parameters for the generic soft-core
potential (Fig.~\ref{potential2}) considered in this paper: $w_R/a$
and $w_A/a$ are the soft-core width and the attractive width,
respectively, both in units of the hard-core distance, and $U_R/U_A$
is the repulsive energy in units of the attractive energy. Sets i-vi
have same $w_A$ and $U_R$; sets ii, vii-xii have same $w_R$ and $U_R$;
sets xii-xvi have same $w_R$ and $w_A$. }
\begin{ruledtabular}
\begin{tabular}{llll}
Set\footnote{The asterisk ($\ast$) denotes sets for which critical points
are calculated via two methods (see Tables~\protect\ref{new}
and \protect\ref{CPs}).}
& $w_R/a$ & $w_A/a$ & $U_R/U_A$ \\
\hline
i     & 0.4     & 0.7   & 2 \\
ii*   & 0.5     & 0.7   & 2 \\
iii*  & 0.6     & 0.7   & 2 \\
iv*   & 0.7     & 0.7   & 2 \\
v     & 0.8     & 0.7   & 2 \\
vi    & 0.9     & 0.7   & 2 \\
vii   & 0.5     & 0.3   & 2 \\
viii  & 0.5     & 0.4   & 2 \\
ix    & 0.5     & 0.5   & 2 \\
x     & 0.5     & 0.6   & 2 \\
xi*   & 0.5     & 0.8   & 2 \\
xii*  & 0.5     & 0.9   & 2 \\
xiii  & 0.5     & 0.9   & 2.5 \\
xiv*  & 0.5     & 0.9   & 3   \\
xv    & 0.5     & 0.9   & 3.5 \\
xvi   & 0.5     & 0.9   & 4   \\
\end{tabular}
\end{ruledtabular}
\end{table}

\begin{table}[hp]
\caption{\label{new} Temperatures $T_{C_1}$ and $T_{C_2}$, pressures
$P_{C_1}$ and $P_{C_2}$, and densities $\rho_{C_1}$ and $\rho_{C_2}$,
for the critical points $C_1$ and $C_2$, respectively, computed by the
isothermic method.}
\begin{ruledtabular}
\begin{tabular}{lrrrrrr}
Set & $k_BT_{C_1}/U_A$ & $a^3P_{C_1}/U_A$ &
    $a^3\rho_{C_1}$
    & $k_BT_{C_2}/U_A$ & $a^3P_{C_2}/U_A$ &
    $a^3\rho_{C_2}$ \\
\hline
ii   &$1.30\pm0.01$ &$0.04\pm0.01$ &$0.11\pm0.02$
        &$0.58\pm0.02$ &$0.15\pm0.02$  &$0.33\pm0.02$\\
iii  &$1.24\pm0.01$ &$0.03\pm0.01$  &$0.09\pm0.02$
        &$0.69\pm0.02$ &$0.11\pm0.02$  &$0.28\pm0.02$\\
iv   &$1.18\pm0.03$ &$0.025\pm0.003$  &$0.08\pm0.02$
        &$0.75\pm0.01$ &$0.07\pm0.01$  &$0.24\pm0.02$\\
xi   &$1.52\pm0.01$ &$0.05\pm0.01$  &$0.11\pm0.02$
        &$0.69\pm0.01$ &$-0.11\pm0.01$ &$0.33\pm0.02$\\
xii  &$1.82\pm0.01$ &$0.06\pm0.02$  &$0.12\pm0.02$
        &$0.96\pm0.02$ &$-0.21\pm0.02$ &$0.32\pm0.03$\\
xiv  &$1.59\pm0.01$ &$0.043\pm0.004$  &$0.10\pm0.02$
        &$0.58\pm0.01$ &$-0.01\pm0.01$ &$0.35\pm0.02$\\
\end{tabular}
\end{ruledtabular}
\end{table}

\begin{table}[hp]
\caption{\label{CPs}
Temperatures $T_{C_1}$ and $T_{C_2}$, pressures $P_{C_1}$ and
$P_{C_2}$, and densities $\rho_{C_1}$ and $\rho_{C_2}$ for the critical
points $C_1$ and $C_2$, respectively, estimated by cooling the
system at constant $\rho$ (isochoric method)  for the potential with the set of
parameters in Table~\protect\ref{sets}.}
\begin{ruledtabular}
\begin{tabular}{lrrrrrr}
Set & $k_BT_{C_1}/U_A$ & $a^3P_{C_1}/U_A$ &
    $a^3\rho_{C_1}$
    & $k_BT_{C_2}/U_A$ & $a^3P_{C_2}/U_A$ &
    $a^3\rho_{C_2}$ \\
\hline
i   &$1.34\pm0.02$ &$0.04\pm0.01$ &$0.13\pm0.02$
       &$0.47\pm0.01$ &$0.28\pm0.01$  &$0.42\pm0.03$\\
ii  &$1.32\pm0.01$ &$0.04\pm0.02$ &$0.11\pm0.02$
       &$0.62\pm0.02$ &$0.19\pm0.02$  &$0.33\pm0.02$\\
iii &$1.25\pm0.01$ &$0.03\pm0.01$ &$0.09\pm0.01$
       &$0.69\pm0.02$ &$0.11\pm0.01$  &$0.29\pm0.02$\\
iv  &$1.19\pm0.01$ &$0.03\pm0.01$  &$0.08\pm0.01$
       &$0.74\pm0.01$ &$0.07\pm0.01$  &$0.26\pm0.02$\\
v   &$1.15\pm0.02$ &$0.02\pm0.02$  &$0.07\pm0.01$
       &$0.75\pm0.01$ &$0.04\pm0.01$  &$0.22\pm0.02$\\
vi  &$1.11\pm0.02$ &$0.02\pm0.02$  &$0.07\pm0.01$
       &$0.76\pm0.01$ &$0.03\pm0.01$  &$0.20\pm0.02$\\ 
vii &$0.68\pm0.01$ &$0.02\pm0.01$  &$0.12\pm0.01$
       &$0.48\pm0.03$ &$2.22\pm0.02$  &$0.46\pm0.06$\\
viii&$0.82\pm0.01$ &$0.03\pm0.01$  &$0.12\pm0.01$
       &$0.52\pm0.03$ &$1.65\pm0.02$  &$0.42\pm0.03$\\
ix  &$0.96\pm0.01$ &$0.03\pm0.02$  &$0.11\pm0.01$
       &$0.53\pm0.03$  &$1.05\pm0.03$  &$0.39\pm0.05$\\
x   &$1.12\pm0.01$ &$0.04\pm0.01$  &$0.10\pm0.01$
       &$0.57\pm0.01$  &$0.58\pm0.01$  &$0.35\pm0.02$\\
xi  &$1.54\pm0.02$ &$0.05\pm0.02$  &$0.12\pm0.01$
       &$0.70\pm0.01$  &$-0.09\pm0.01$ &$0.33\pm0.03$\\  
xii &$1.84\pm0.02$ &$0.06\pm0.02$  &$0.13\pm0.01$
       &$0.96\pm0.01$  &$-0.22\pm0.01$ &$0.31\pm0.03$\\
xiii&$1.67\pm0.01$ &$0.05\pm0.01$  &$0.11\pm0.01$
       &$0.72\pm0.01$  &$-0.15\pm0.01$ &$0.35\pm0.01$\\
xiv &$1.62\pm0.02$ &$0.05\pm0.01$  &$0.09\pm0.01$
       &$0.60\pm0.01$  &$0.01\pm0.01$  &$0.37\pm0.04$\\
xv  &$1.57\pm0.01$ &$0.04\pm0.01$  &$0.09\pm0.01$
       &$0.55\pm0.01$  &$0.28\pm0.01$  &$0.35\pm0.02$\\
xvi &$1.54\pm0.01$ &$0.04\pm0.01$  &$0.09\pm0.01$
       &$0.53\pm0.02$  &$0.60\pm0.02$  &$0.35\pm0.02$\\
\end{tabular}
\end{ruledtabular}
\end{table}

\begin{table}[hp]
\caption{\label{summ-table} 
Summary of the effects on $\rho_{C_1}$, $T_{C_1}$,
$P_{C_1}$, and $\rho_{C_2}$, $T_{C_2}$,
$P_{C_2}$, from variation of parameters $w_A/a$, $w_R/a$, and
$U_R$, one at the time. The symbols $\uparrow$, $\downarrow$ and
$\approx$ represent, respectively, an increase, a decrease and a small
variation of a thermodynamic quantity as a consequence of the increase
of the potential parameter.}
\begin{ruledtabular}
\begin{tabular}{c|cccccc}
 & $\rho_{C_1}$ & $T_{C_1}$    & $P_{C_1}$ & $\rho_{C_2}$ &
  $T_{C_2}$  & $P_{C_2}$ \\
\hline
$w_A/a$     & $\approx$    & $\uparrow$   & $\uparrow$ & $\downarrow$  &
  $\uparrow$ & $\downarrow$ \\
$w_R/a$     & $\downarrow$ & $\downarrow$ & $\downarrow$ & $\downarrow$
  & $\uparrow$ & $\downarrow$ \\
$U_R/U_A$ & $\downarrow$ & $\downarrow$ & $\downarrow$ & $\approx$ &
  $\downarrow$  & $\uparrow$\\
\end{tabular}
\end{ruledtabular}
\end{table}

\begin{table}[hp]
\caption{\label{CP-single-well}
Density $\rho_c$, temperature $T_c$, and pressure $P_c$
for the gas-liquid critical point for a single square-well potential
with attractive range $w_A$, attractive energy $U_A$ and hard-core
distance $a$.}
\begin{ruledtabular}
\begin{tabular}{cccc}
$w_A/a$  & $a^3\rho_c$     & $k_BT_c/U_A$    & $a^3P_c/U_A$ \\
\hline
$0.1$    & $0.45\pm0.05$ & $0.48\pm0.01$  & $0.062\pm0.005$ \\
$0.2$    & $0.42\pm0.05$ & $0.67\pm0.01$  & $0.076\pm0.005$ \\
$0.3$    & $0.34\pm0.05$ & $0.86\pm0.01$  & $0.088\pm0.005$ \\
$0.4$    & $0.35\pm0.05$ & $1.04\pm0.01$  & $0.094\pm005$ \\
$0.5$    & $0.30\pm0.03$ & $1.24\pm0.01$  & $0.103\pm003$ \\
$0.6$    & $0.28\pm0.03$ & $1.45\pm0.01$  & $0.113\pm003$ \\
\end{tabular}
\end{ruledtabular}
\end{table}

\begin{table}
\caption{Single square-well potential: parameters of the linear fits in
Fig.~\protect\ref{fit-single-well} for
$k_BT_c/U_A=p + q w_A/a$, and analogous linear expressions for
$a^3P_c/U_A$ and $a^3\rho_c$, for the gas-liquid critical point.}
\begin{ruledtabular}
\begin{tabular}{c|c c c}
          & $k_BT/U_A$     & $a^3P/U_A$        & $a^3\rho$         \\
\hline
$p$  & $0.29 \pm 0.01$& $0.055 \pm 0.002$ & $0.482 \pm 0.008$ \\
$q$  & $1.91 \pm 0.03$& $0.097 \pm 0.006$ & $-0.35 \pm 0.02$
\end{tabular}
\end{ruledtabular}
\label{fit-parameters-single-well}
\end{table}


\break\newpage


\begin{center}
\begin{figure}[htb]
\centering
\includegraphics[width=8.0cm,height=10.0cm,angle=-90]{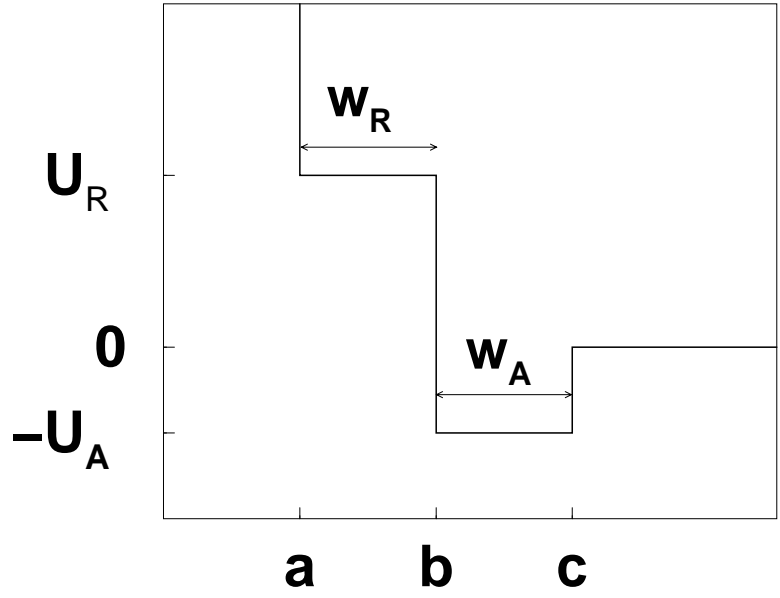}
\caption{ The generic soft-core potential with attractive well with
parameters $w_A/a$, $w_{R}/a$, and $U_{R}/U_A$. We use the parameters
listed in Table~\protect\ref{sets}.}
\label{potential2}
\end{figure}
\end{center}

\begin{center}
\begin{figure}[ht]
\centering
\includegraphics[width=10.0cm,height=12.0cm,angle=-90]{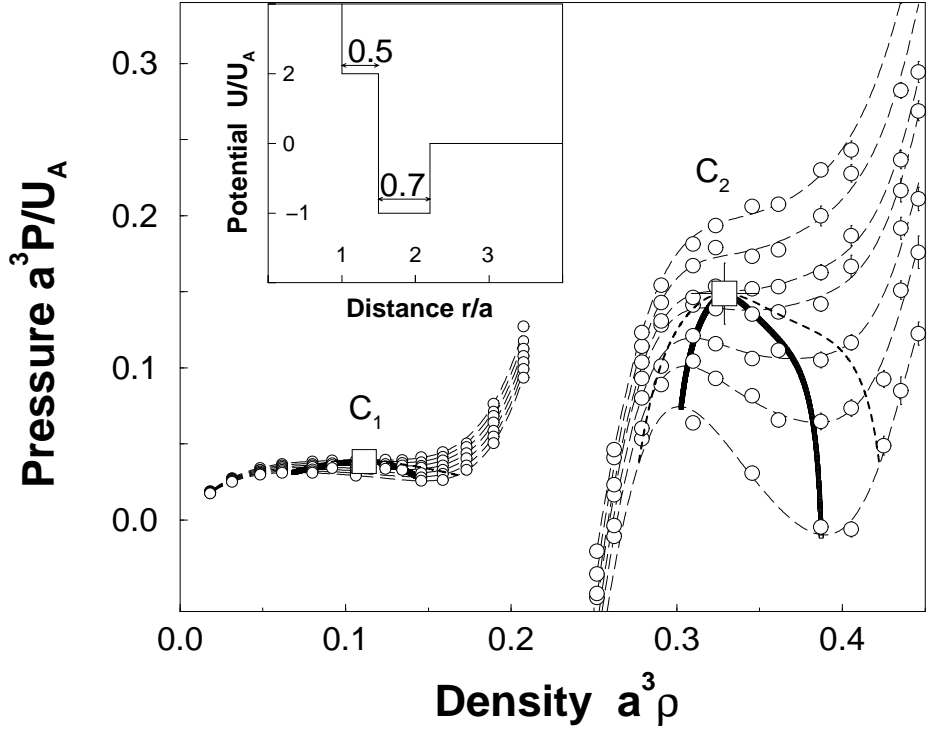}
\caption{The MD $P$-$\rho$ phase diagram for the potential in the inset,
  with the parameter set ii in Table~\ref{sets}. The long-dashed lines
  are the fits of the calculated state points (circles) at constant
  $T$. The isotherms (from top to bottom) are for $k_BT/U_A=1.30$, 1.29,
  1.28, 1.27, 1.26, 1.25, 1.24 at low-$\rho$ and $k_BT/U_A=0.62$, 0.60,
  0.58, 0.57, 0.55, 0.53, 0.50 at high-$\rho$. The fits are calculated
  by considering $P$ a polynomial function of both $T$ and $\rho$. The
  isotherms show two regions with negative slope, i.e. mechanically
  unstable, delimited by the spinodal lines (solid bold lines). Each
  spinodal line is associated with a first-order phase transition.  By
  using the Maxwell construction, we estimate the coexisting regions
  associated to each spinodal line, delimited by the phase transition
  line (bold dashed line).  The coexisting regions are clearly separated
  at the considered temperatures. The phase transition line at low
  $\rho$ is indistinguishable from the spinodal line at this scale.  The
  points where the coexisting lines merge with the spinodal lines are,
  by definition, the critical points $C_1$ (at low-$\rho$) and $C_2$ (at
  high-$\rho$).  No spontaneous crystal nucleation is observed in the explored
  region of the phase diagram. }
\label{md1}
\end{figure}
\end{center}

\begin{center}
\begin{figure}[ht]
\centering
\includegraphics[width=10.0cm,height=12.0cm,angle=-90]{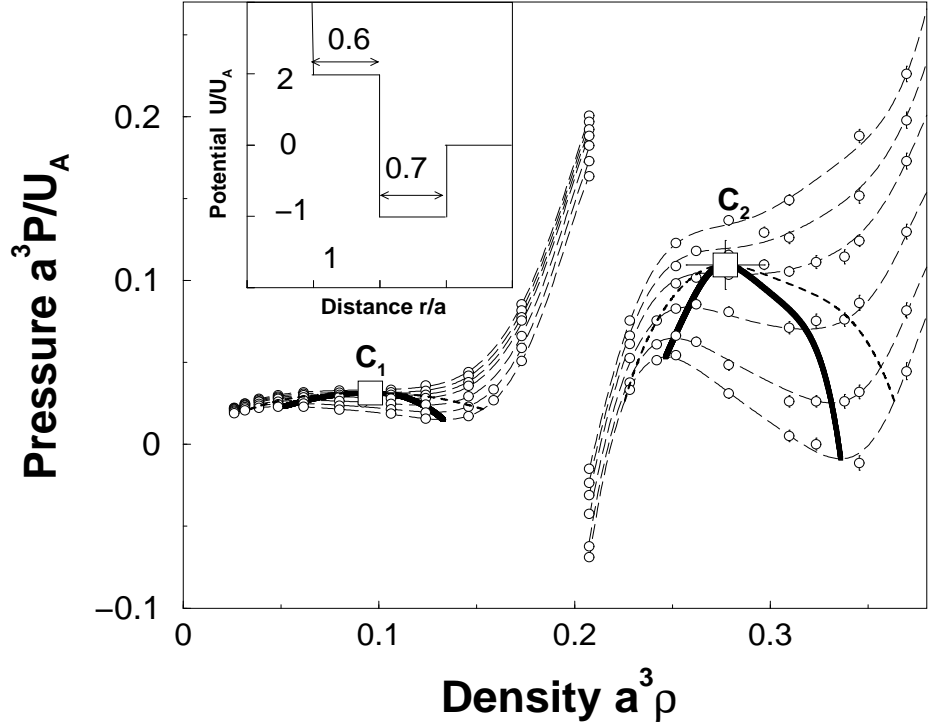}
\caption{As in Fig.~\ref{md1}, for parameter set iii in Table~\ref{sets}.
The isotherms in the low-$\rho$ region  (from top to bottom) are for
$k_BT/U_A=1.25$, 1.24, 1.23, 1.22, 1.20, 1.18, 1.16, and in the
high-$\rho$ region are for $k_BT/U_A=0.72$, 0.70, 0.68, 0.65, 0.62,
0.60.  Spontaneous crystal nucleation is observed for $T<T_{C_2}$ and
$\rho>\rho_{C_2}$.}
\label{md2}
\end{figure}
\end{center}

\begin{center}
\begin{figure}[ht]
\centering
\includegraphics[width=12.0cm,height=10.0cm,angle=0]{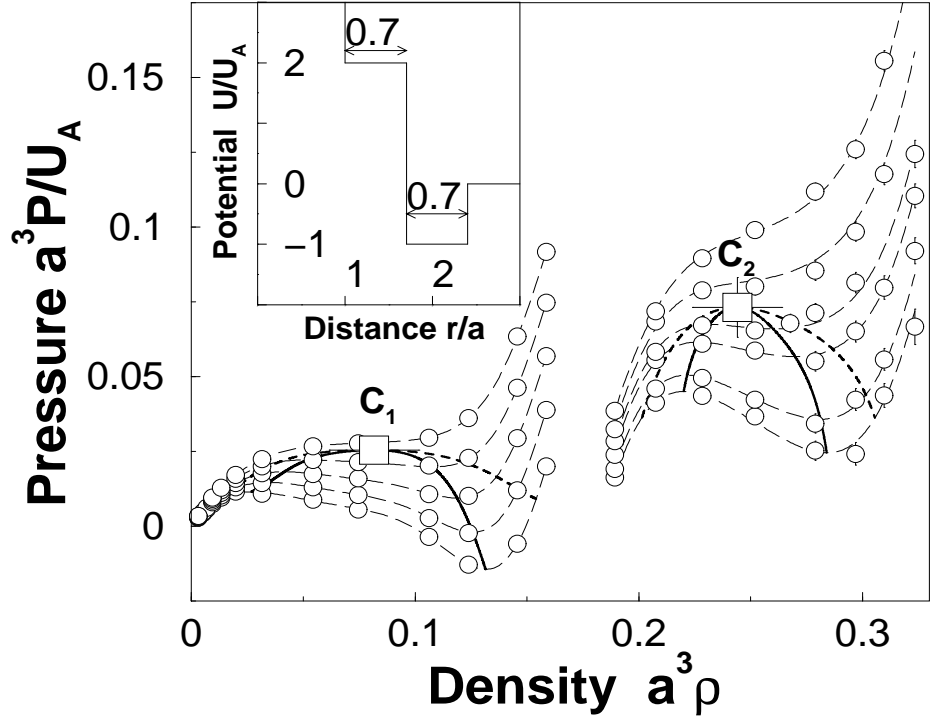}
\caption{ As in Fig.~\ref{md1}, for parameter set iv in
Table~\ref{sets}.  The isotherms (from top to bottom) in the low-$\rho$
region are for $k_BT/U_A=1.20$, 1.15, 1.10, 1.05, 1.00, and in the
high-$\rho$ region are for $k_BT/U_A=0.77$, 0.75, 0.73, 0.72, 0.70,
0.69.  No spontaneous crystal nucleation is observed in the explored region of the
phase diagram. }
\label{md3}
\end{figure}
\end{center}

\begin{center}
\begin{figure}[ht]
\centering
\includegraphics[width=10.0cm,height=12.0cm,angle=-90]{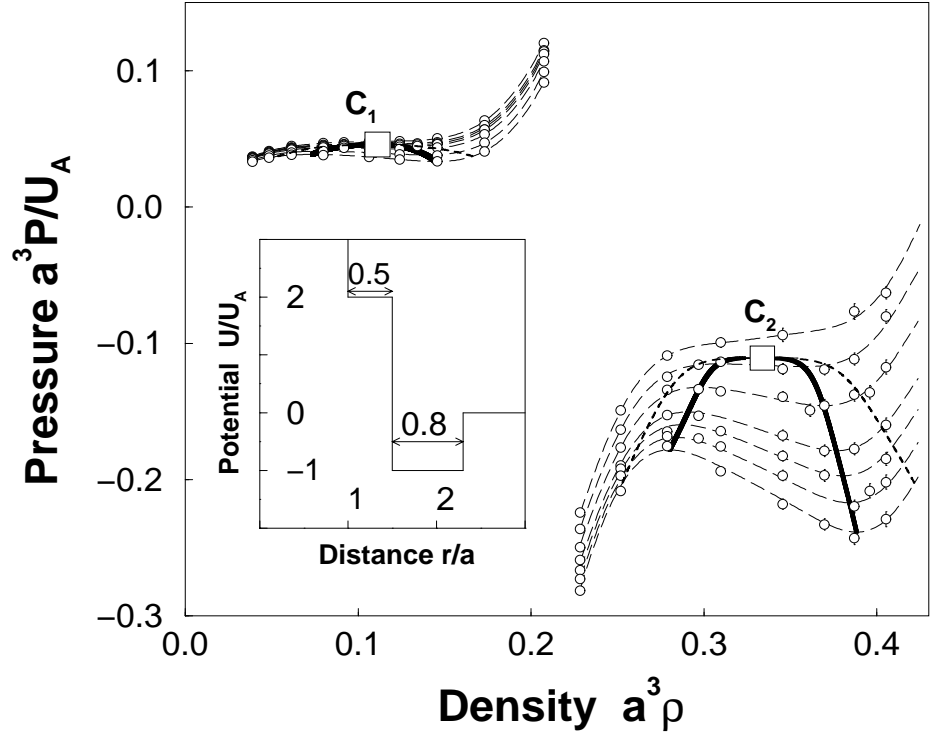}
\caption{ As in Fig.~\ref{md1}, for parameters xi in Table~\ref{sets}.
The isotherms (from top to bottom) in the low-$\rho$ region are for
$k_BT/U_A=1.53$, 1.52, 1.515, 1.51, 1.50, 1.48, 1.46, and in the
high-$\rho$ region are for $k_BT/U_A=0.70$, 0.68, 0.66, 0.64, 0.63,
0.62, 0.61. $C_2$ is at negative pressure.}
\label{md4}
\end{figure}
\end{center}

\begin{center}
\begin{figure}[ht]
\centering
\includegraphics[width=12.0cm,height=10.0cm,angle=0]{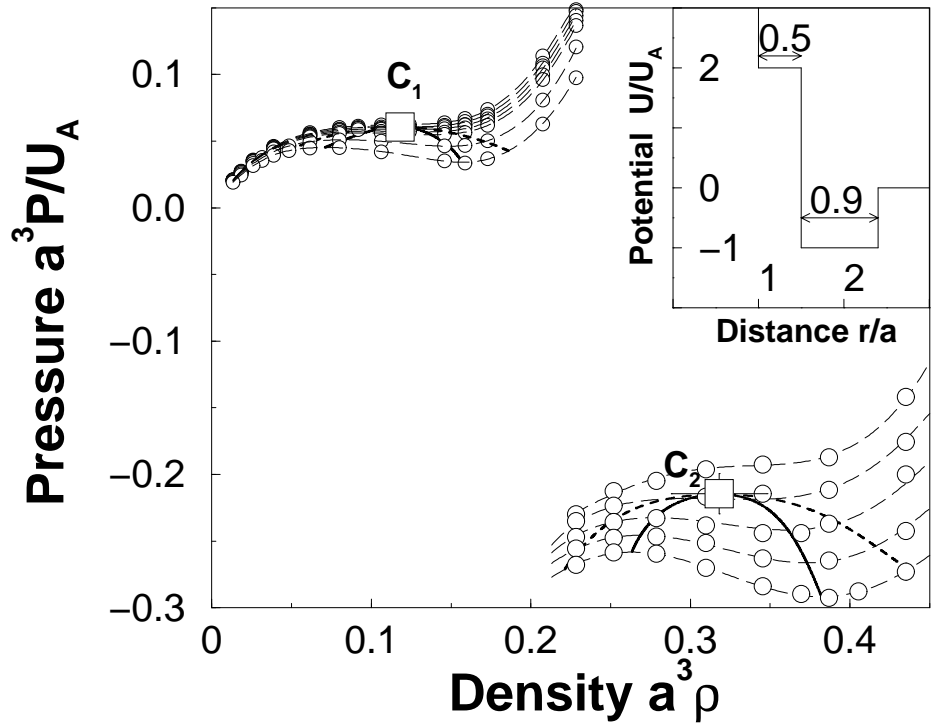}
\caption{ As in Fig.~\ref{md1}, for parameters xii in Table~\ref{sets}.
The isotherms (from top to bottom) in the low-$\rho$ region are for
$k_BT/U_A=1.83$, 1.82, 1.815, 1.81, 1.80, 1.79, 1.75, 1.70, and in the
high-$\rho$ region are for $k_BT/U_A=0.98$, 0.96, 0.64, 0.92,
0.90. $C_2$ is at a negative pressure.}
\label{md5}
\end{figure}
\end{center}

\begin{center}
\begin{figure}[ht]
\centering
\includegraphics[width=10.0cm,height=12.0cm,angle=-90]{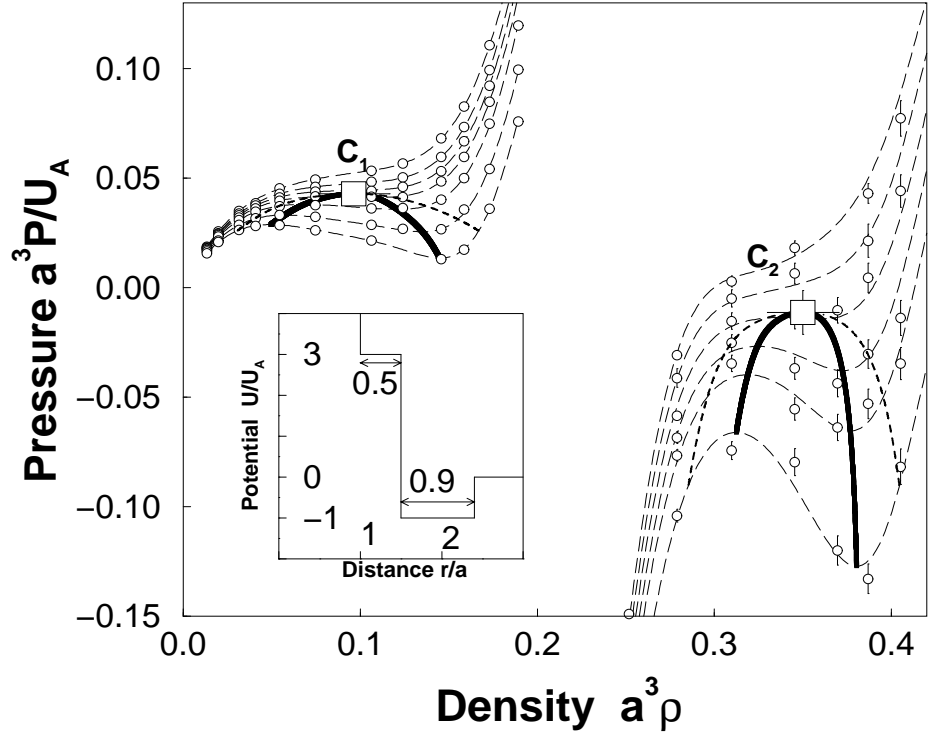}
\caption{ As in Fig.~\ref{md1}, for parameters xiv in
Table~\ref{sets}.  The isotherms (from
top to bottom) 
in the low-$\rho$ region are for 
$k_BT/U_A=1.65$, 1.62, 1.60, 1.58, 1.55, 1.50, 1.45, and
in the high-$\rho$ region are for $k_BT/U_A=0.60$, 0.59, 0.58, 0.57,
0.56, 0.54. $C_2$ is at negative pressures.}
\label{md6}
\end{figure}
\end{center}

\begin{center}
\begin{figure}[h]
\centering
\includegraphics[width=9.0cm,height=10.8cm,angle=-90]{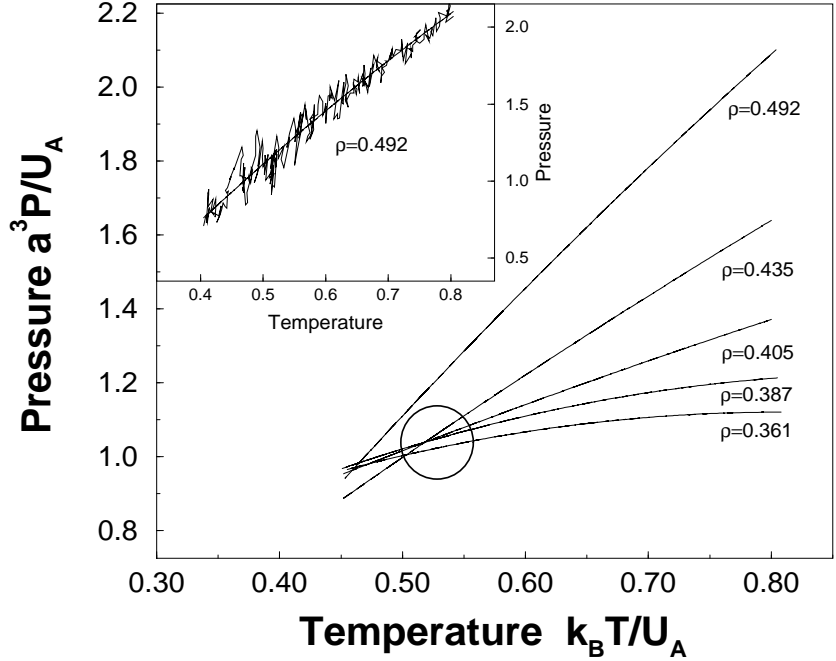}
\caption{Estimation of the critical point $C_2$ by the isochoric method
for the set of potential parameters $w_A/a=0.5$, $w_R/a=0.5$,
$U_R/U_A=2$.  Inset: $P$ at constant $a^3\rho=0.492$ for the MD
calculation during the slow cooling described in the text. For
$k_BT/U_A>0.5$ the errors on the estimate of the state points are of the
order of the non-monotonic jumps. The interpolating line is a quadratic
fit of the calculated points, and gives an estimate of the isochore at
$a^3\rho=0.492$ for $k_BT/U_A>0.5$.  Main panel: quadratic fits of
isochores for $a^3\rho= 0.492$, 0.435, 0.405, 0.387, 0.361 (from top to
bottom). The critical point $C_2$ is located at the highest-$T$
intersection of two isochores (region inside the circle). The
indeterminacy of this intersection gives an estimate of the error on
the values $T_{C_2}=0.53\pm 0.03$, $P_{C_2}=1.05\pm 0.03$, and
$\rho_{C_2}=0.39\pm 0.05$.}
\label{isochores}
\end{figure}
\end{center}

\begin{center}
\begin{figure}[ht]
\centering
\includegraphics[width=16.0cm,height=14.0cm,angle=0]{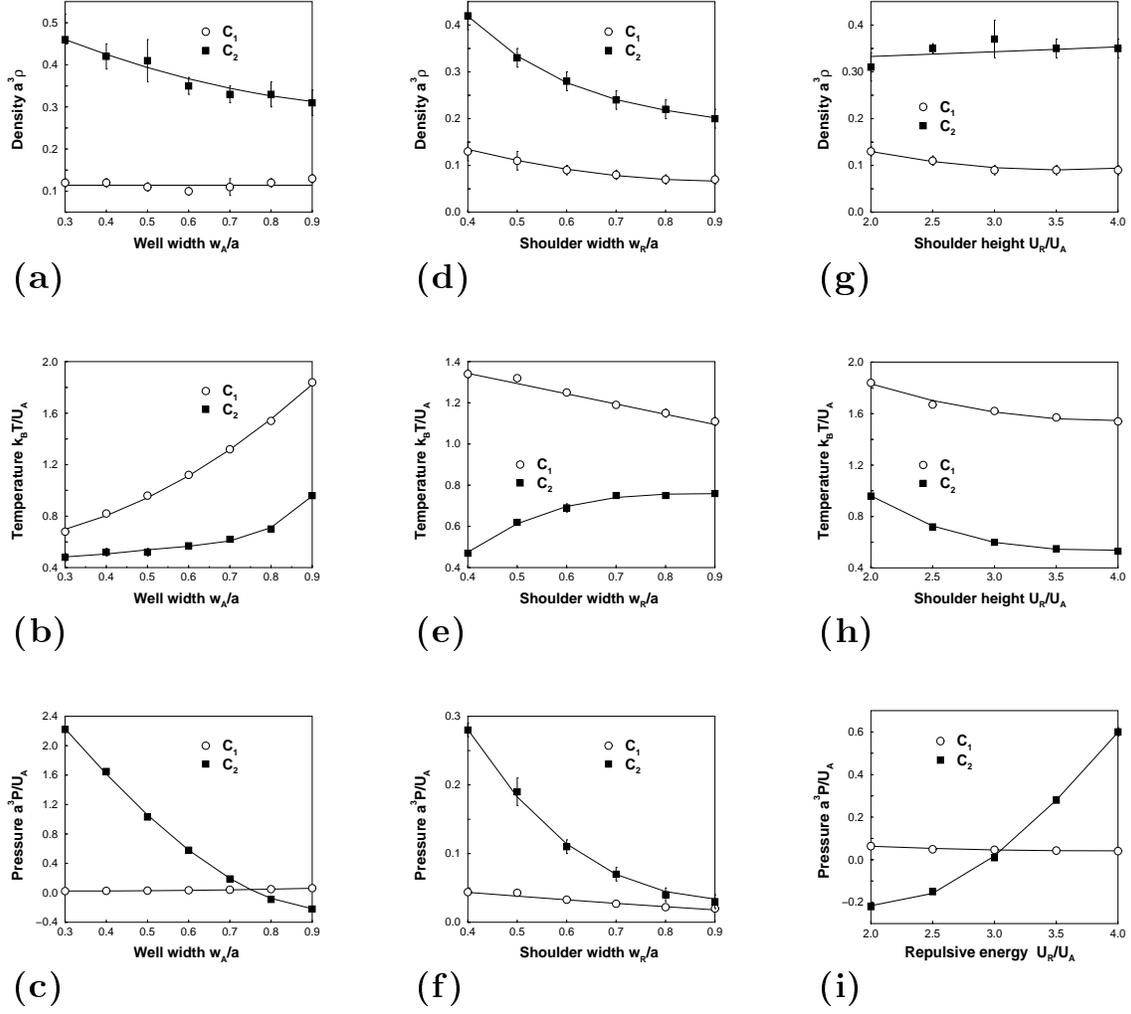}\\
\vspace{3.0cm}
\caption{The behavior of the density, temperature and pressure of the
  low-density critical point $C_1$ (open circles) and high-density
  critical point $C_2$ (filled squares) for variations of the potential
  parameters (a)-(c) $w_{A}$, (d)-(f) $w_{R}$ and (g)-(i) $U_{R}$. The
  other two parameters are constant: in (a)-(c) $w_{R}/a=0.5$ and
  $U_{R}/U_{A}=2$, in (d)-(f) $w_{A}/a=0.7$ and $U_{R}/U_{A}=2$, in
  (g)-(i) $w_{A}/a=0.9$ and $w_{R}/a=0.5$.  Where not shown, errors are
  smaller than the symbol size.  Lines are guides for the eye.}
\label{all}
\end{figure}
\end{center}

\begin{center}
\begin{figure}[ht]
\centering
\includegraphics[width=6.5cm,height=7.5cm,angle=-90]{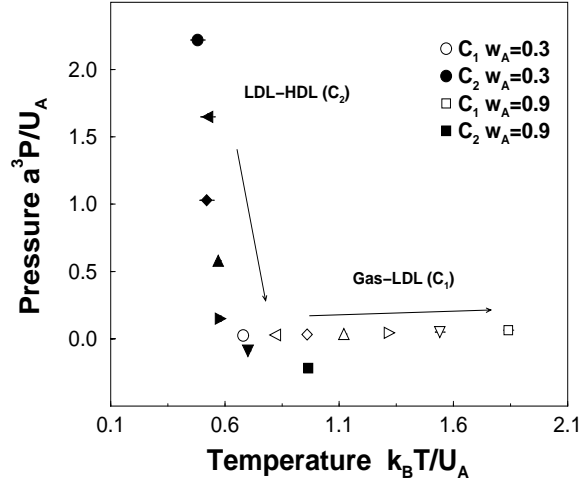}\\
\caption{The gas-LDL critical point ($C_1$) and LDL-HDL critical point
  ($C_2$) in the $P-T$ plane, for varying attractive width $w_{A}$ and
  constant $w_{R}/a=0.5$, $U_{R}/U_A=2.0$. Symbols denote: 
  $w_{A}/a=0.3$ (circles), 0.4 (left triangles), 0.5 (diamonds), 0.6 (up
  triangles), 0.7 (right triangles), 0.8 (down triangles), 0.9
  (squares). Open symbols are for $C_1$ and filled symbols are for
  $C_2$. The arrows denote the direction of increasing $w_A$.}
\label{wl_w12.gc}
\end{figure}
\end{center}

\begin{center}
\begin{figure}[ht]
\centering
\includegraphics[width=6.5cm,height=7.5cm,angle=-90]{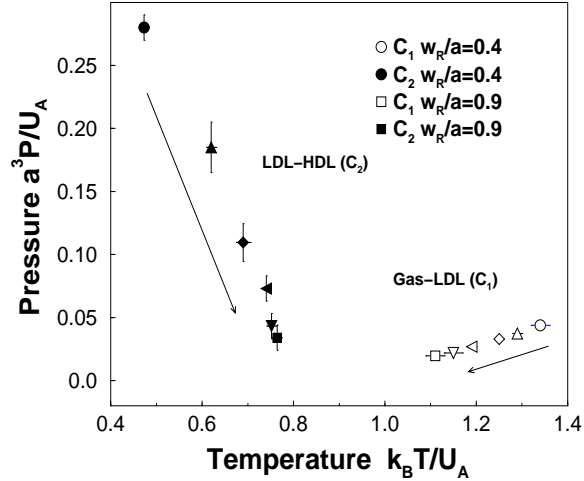}\\
\caption{ The gas-LDL critical point ($C_1$) and LDL-HDL critical point
($C_2$) in the $P-T$ plane, for varying shoulder width $w_{R}/a$ and
constant $w_{A}/a=0.7$, $U_{R}/U_A=2.0$. Symbols denote: $w_{R}/a=0.4$
(circles), 0.5 (up triangles), 0.6 (diamonds), 0.7 (left triangles), 0.8
(down triangles), 0.9 (squares). Open symbols are for $C_1$ and filled
symbols are for $C_2$. The arrows denote the direction of increasing
$w_R$.}
\label{sh_w12.gc}
\end{figure}
\end{center}

\begin{center}
\begin{figure}[ht]
\centering
\includegraphics[width=6.5cm,height=7.5cm,angle=-90]{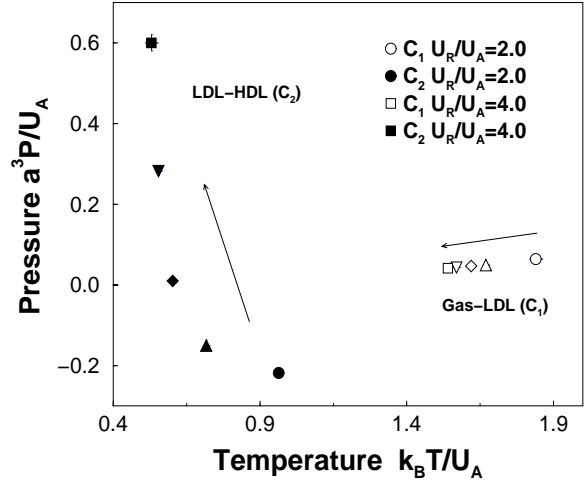}\\
\caption{The gas-LDL critical point ($C_1$) and LDL-HDL critical point
($C_2$) in the $P-T$ plane, for varying repulsive energy $U_{R}/U_A$
and constant $w_{A}/a=0.9$, $w_{R}/a=0.5$. Symbols denote:
$U_{R}/U_A=2.0$ (circles), 2.5 (up triangles), 3.0 (diamonds), 3.5 (down
triangles), 4.0 (squares).  Open symbols are for $C_1$ and filled
symbols are for $C_2$. The arrows denote the direction of increasing
$U_R$.}
\label{sh_h12.gc}
\end{figure}
\end{center}

\begin{center}
\begin{figure}[ht]
\centering
\includegraphics[width=7.0cm,height=8.0cm,angle=270]{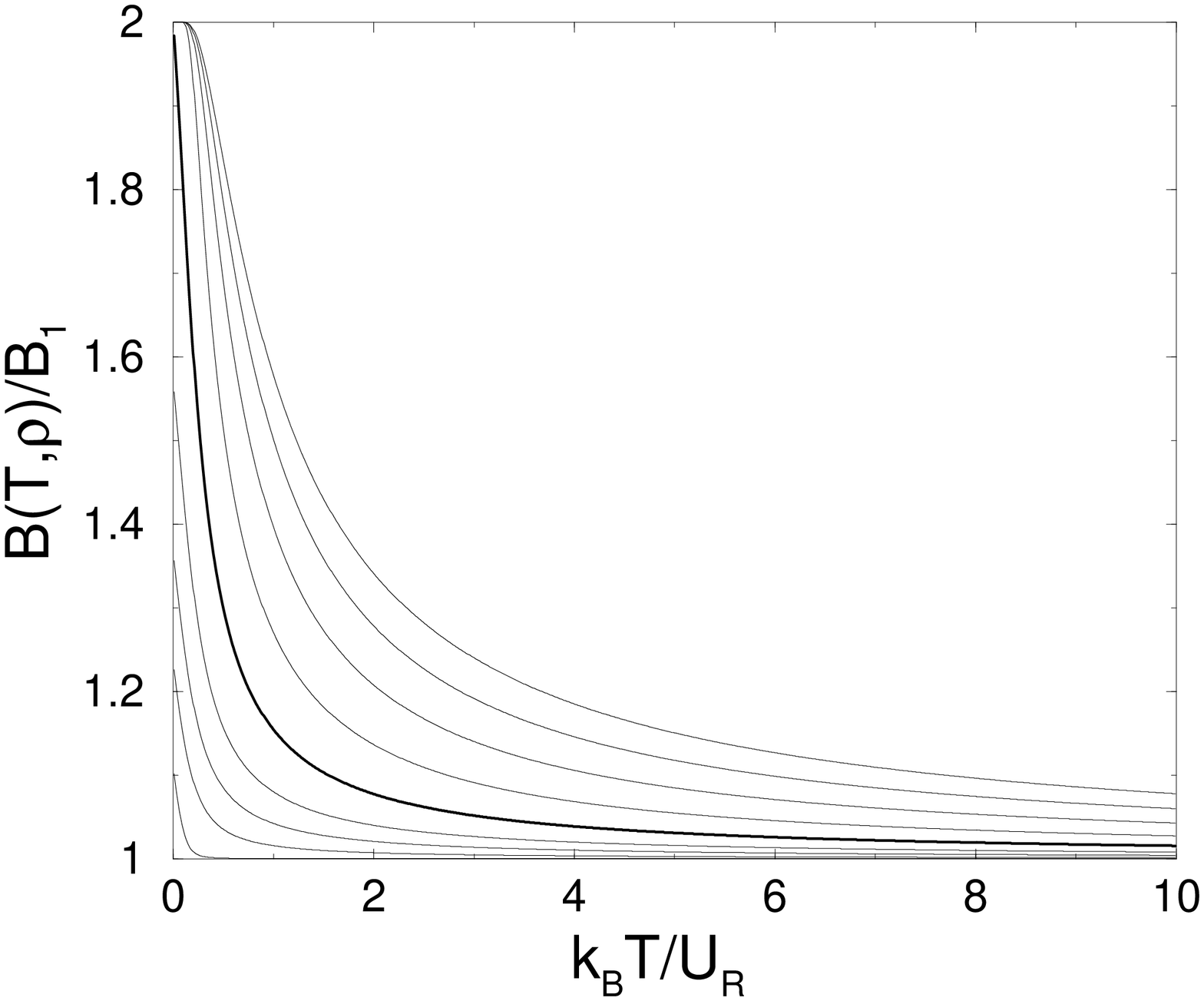}
\includegraphics[width=7.0cm,height=8.0cm,angle=270]{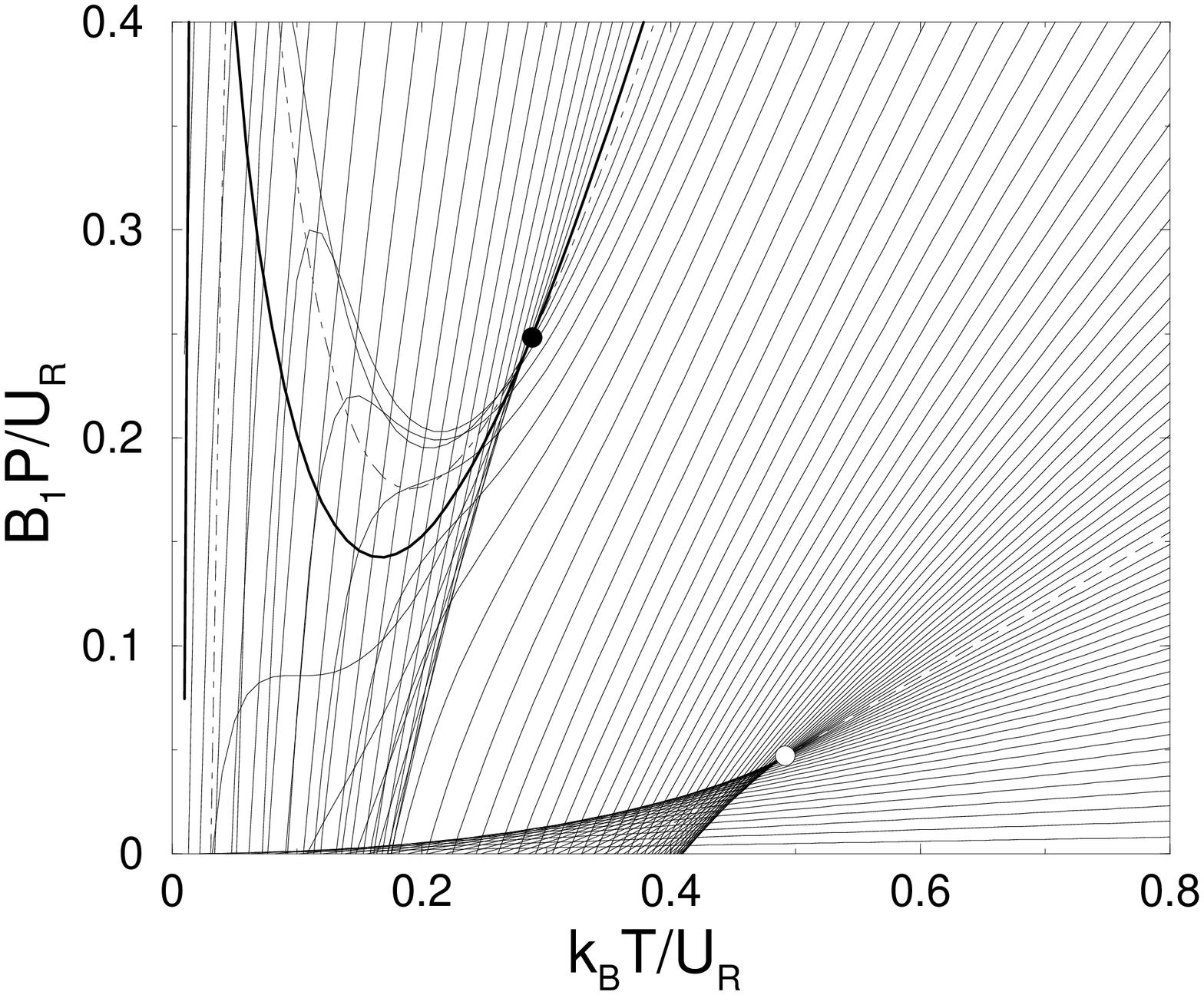}
\caption{(a) The behavior of the effective excluded volume 
$B(\rho,T)$
for a one dimensional system with $B_2/B_1=2$ for
densities $B_1\rho=0.1,0.2,0.3,0.4,0.5,0.6,0.7,0.8,0.9$ from top to
bottom. The thick curve indicates the behavior for
$B_1\rho=0.5=B_1/B_2$. (b) The isochores of MVDWE for $B_2/B_1=1.4$,
$A/(B_1U_R)=2.2$ on the $P-T$ plane. An open circle
indicates the low density critical point. A filled circle indicates
high-density critical point.  A dashed line indicates the low-density
critical isochore $B_1\rho_{C_1}\approx 0.25$.  A dot-dashed line indicates the
high-density critical isochore $B_1\rho_{C_2} \approx 0.70$. The thick line
indicates $B_1\rho \approx 0.71=B_1/B_2$. Note that isochores start to develop
density anomaly below the high-density critical point. }
\label{vander}
\end{figure}
\end{center}

\begin{center}
\begin{figure}[ht]
\centering
\includegraphics[width=16.0cm,height=14.0cm,angle=0]{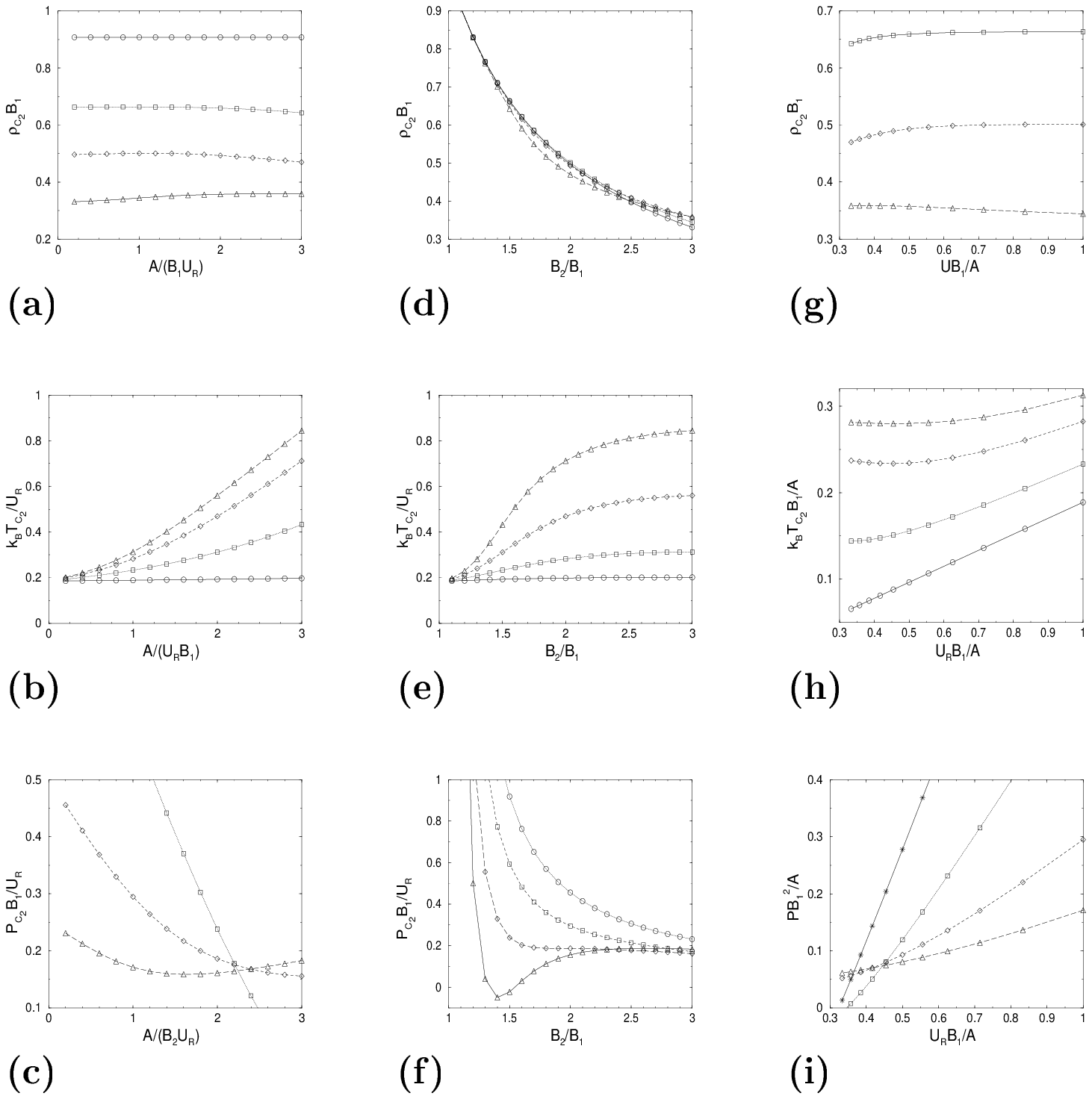}\\
\vspace{3.0cm}
\caption{The behavior of the density, temperature and pressure of the
high-density
  critical point $C_2$ for variations of the MVDWE
  parameters (a)-(c) $A$, (d)-(f) $B_2$ and (g)-(i) $U_{R}$. The
  other two parameters are constant: 
in (a)-(c) $B_2/B_1=1.1~(\circ ),~1.5~(\Box ),~2.0~(\diamond ),~3.0~(\triangle )$ and  $U_{R}/U_{A}=1.0$, 
in (d)-(f) $A/(B_1U_R)=0.2~(\circ ),~1.0~(\Box ),~2.0~(\diamond ),~3.0~(\triangle )$ and $U_{R}/U_{A}=1$, 
in (g)-(i) $A/(B_1U_R)=1.0$ and $B_2/B_1=1.1~(\circ ),~1.3~(\ast ),~1.5~(\Box ),~2.0~(\diamond ),~3.0~(\triangle )$.
Lines are guides for the eye.}
\label{new3x3}
\end{figure}
\end{center}

\begin{center}
\begin{figure}[ht]
\centering
\includegraphics[width=8.0cm,height=7.0cm,angle=0]{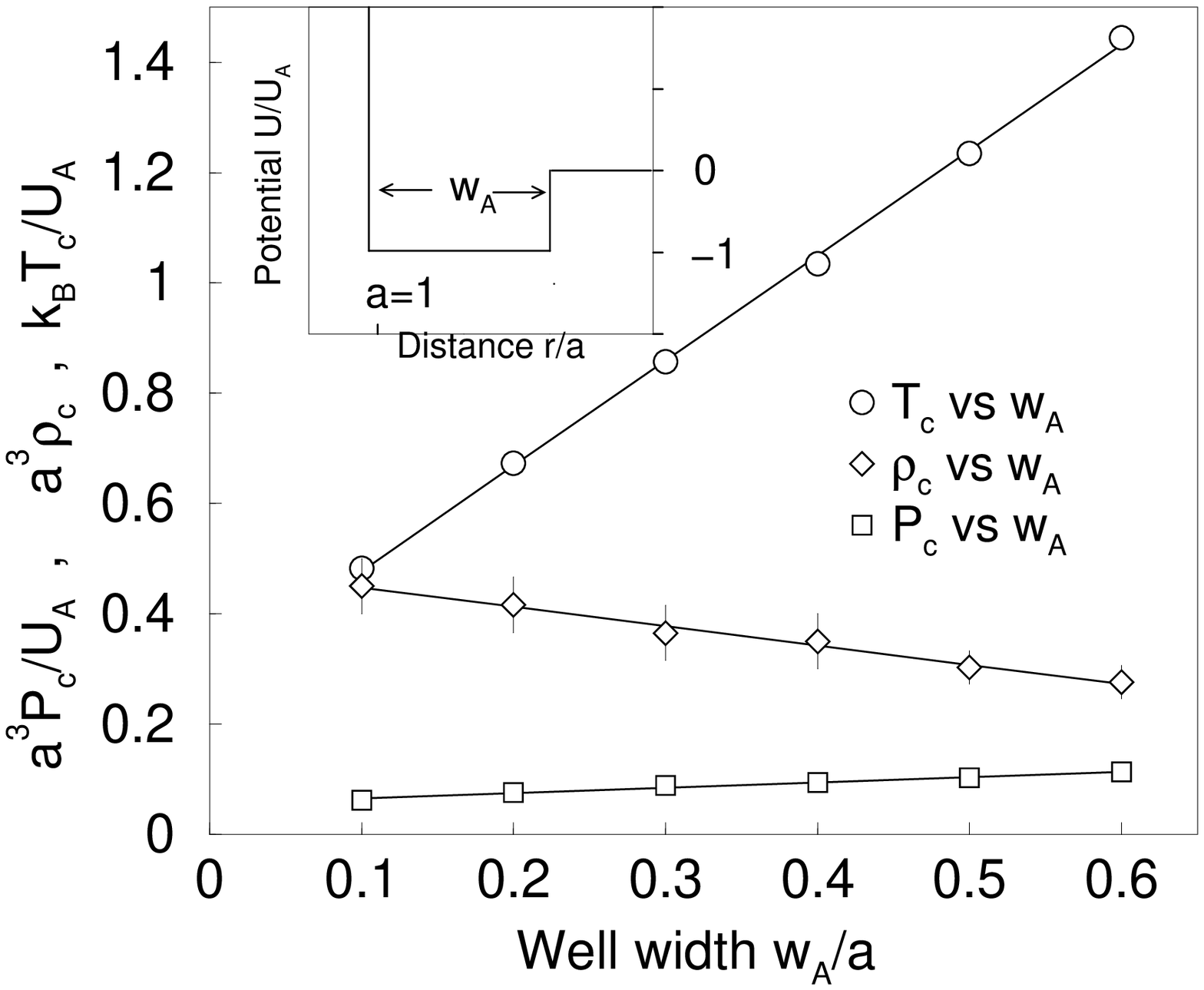}
\caption{Inset: the single square-well potential defined by the well
width $w_A$. Main panel: symbols represent the values of the normalized
temperature $T_c$ (circles), the density $\rho_c$ (diamonds), and
the pressure $P_c$ (squares)
of the gas-liquid critical point for
different values of the parameter $w_A$.  Where not shown, errors are smaller
than the symbol size.  Lines are the linear fits $T_c$, $P_c$, and
$\rho_c$ as functions of $w_A$, with the parameters in
Table~\protect\ref{fit-parameters-single-well}.}
\label{fit-single-well}
\end{figure}
\end{center}


\begin{thebibliography}{99}
\bibitem{ENERGY} To gradually decrease the temperature of the system
from $T_I$ to $T_F$, we rescale the velocities of all the particles
after each $N$ collisions of particles. The velocity of each particle is
multiplied by $\sqrt{1-(1-E_\tau/E_F)\kappa \tau}$, where
$E_F=(3/2)k_BT_F$ is the average kinetic energy of the particles at
$T_F$, $E_\tau$ is the kinetic energy of the particles averaged over
time interval $\tau$ during which $N$ collisions happen, and
$\kappa=5\cdot10^{-4}$ is the heat exchange coefficient, that has the
dimension of the inverse time unit.
\end{thebibliography}
\end{document}